\documentstyle[aps,prl,epsfig,floats]{revtex}

\def\gsim{\lower0.5ex\hbox{$\:\buildrel >\over\sim\:$}}
\def\lsim{\lower0.5ex\hbox{$\:\buildrel <\over\sim\:$}}
\def \rp{{R\hspace{-0.22cm}/}_P}
\def \lp{{L\!\!\!/}}
\def \n{\noindent}

\begin{document}

%%\preprint{}

\title{New scalar resonances from sneutrino--Higgs mixing in
supersymmetry with small lepton number (R-parity) violation}
\author{S. Bar-Shalom$^a$, G. Eilam$^b$ and B. Mele$^c$}
\address{$^a$ Theoretical Physics Group, Rafael, Haifa 31021, 
Israel.\footnote{e-mail: shaouly@physics.technion.ac.il}\\
$^b$ Physics Department, Technion-Institute of Technology, 
Haifa 32000, Israel.\footnote{e-mail: eilam@physics.technion.ac.il}\\
$^c$ INFN, Sezione di Roma and Dept. of Physics, 
University of Roma I, La Sapienza, Roma, Italy.\footnote{e-mail: Barbara.Mele@roma1.infn.it}}
\date{\today }
\maketitle

\begin{abstract}
We consider new $s$-channel scalar exchanges in top quark
and massive gauge-bosons pair production in $e^+e^-$ collisions, 
in supersymmetry with a
small lepton number violation. We show that a soft bilinear lepton
number violating term in the scalar potential which mixes the Higgs and
the slepton fields can give rise to a significant scalar resonance
enhancement in $e^+e^- \to ZZ,~W^+W^-$ and in $e^+e^- \to t \bar
t$. The sneutrino--Higgs mixed state couples to the incoming
light leptons through its sneutrino component and to either the top quark
or the massive gauge bosons through its Higgs component. Such
a scalar resonance in these specific production channels cannot
result from trilinear Yukawa-like $R$-parity violation alone, and
may, therefore, stand as strong evidence for the existence of
$R$-parity violating bilinears in the supersymmetric scalar
potential. We use the LEP2 measurements of the $WW$ and $ZZ$
cross-sections to place useful constrains on this scenario, and
investigate the expectations for the sensitivity of a future
linear collider to these signals. We find that signals of
these scalar resonances , in particular in top-pair production,
are well within the reach of linear colliders
in the small lepton number violation scenario.
\end{abstract}

\draft

\section{Introduction}

Despite the enormous success of the Standard Model (SM), the SM
spectrum and dynamics are believed to be the low energy limit of a
more fundamental theory. There are, indeed, strong theoretical
motivations for the existence of new physics above the electroweak
mass scale: the mysterious large hierarchy from the electroweak
scale to the Planck scale, the lack of a theory that unifies
quantum physics with gravity, the observed dark matter in
the universe etc...

One of the most favorable new physics candidates which may provide
a viable framework for such questions is supersymmetry
(SUSY). From the phenomenological point of view, SUSY offers some
new attractive features which are absent in the SM and which may
be tested in upcoming future colliders. One example of a
fundamental difference between SUSY and the SM is associated with
lepton number. In the SM lepton number must be conserved since it
is not possible to write down a renormalizable lepton number
violating interaction out of the SM fields. In SUSY, however, as
opposed to the SM, lepton number does not have to be conserved
since the most general set of SUSY renormalizable operators does
allow for lepton number violating interactions. In particular, if
one does not impose the so called R-parity symmetry
\cite{rpreview} on the SUSY Lagrangian, then lepton number (and
baryon number) can be violated at tree-level in interaction
vertices involving both sparticles and particles.

Therefore, since there is no fundamental principle that enforces
lepton number conservation, it is clear that lepton number
violating phenomena should be explored in collider experiments even
in processes not involving external SUSY partners.
Such searches will provide an unambiguous test of the SM, and
may give a first solid
evidence about SUSY dynamics.

The SUSY R-parity conserving (RPC) superpotential can be written
as (see e.g., \cite{rosiek} and references therein):

\begin{equation}
{\cal W}_{\rm RPC} = \epsilon_{ab} \left[ \frac{1}{2} h_{jk} {\hat
H}^a_d {\hat L}^b_j {\hat E}_k^c + h_{jk}^\prime {\hat H}^a_d
{\hat Q}^b_j {\hat D}_k^c + h_{jk}^{\prime \prime} {\hat H}^a_u
{\hat Q}^b_j {\hat U}_k^c
 - \mu_0 {\hat H}^a_d {\hat H}_u^b \right] ~,
\label{lrpc}
\end{equation}

\noindent where  $\hat H_u(\hat H_d)$ are the up(down)-type Higgs
supermultiplet and ${\hat L}({\hat E}^c)$ are the leptonic SU(2)
doublet(charged singlet) supermultiplets. The ${\hat Q}$ are quark
SU(2) doublet supermultiplets and ${\hat U}^c({\hat D}^c)$ are
SU(2) up(down)-type quark singlet supermultiplets. Also, $j,k
=1,2$ or $3$ are generation labels and $a,b=1,2$ are SU(2)
indices where $\epsilon_{ab}$ is the rank 2 anti-symmetric tensor.

If R-parity is violated, then lepton number may no longer be a
conserved quantum number of the theory. In this case the $\hat L$
and $\hat H_d$ superfields, which have the same gauge quantum
numbers, lose their identity since there is no additional quantum
number that distinguishes between them. One can then construct
additional renormalizable R-parity violating (RPV) interactions
simply by replacing $\hat H_d \to \hat L$ in (\ref{lrpc}). Thus,
the SUSY superpotential can violate lepton number (or more
generally R-parity)\footnote{We will loosely refer to the SUSY
interactions in (\ref{lp}) and (\ref{bterm}) either as lepton
number violating or RPV interactions.} via an RPV Yukawa-like
trilinear term (RPVTT) in the pure leptonic sector and via a
mass-like RPV bilinear term (RPVBT) as follows
\cite{rpreview}:\footnote{The lepton number violating RPVTT
$\lambda^\prime \hat L \hat Q \hat D^c$ is not relevant for our
calculations.}
\begin{equation}
{\cal W}_{\rp,\lp} \supset \epsilon_{ab} \left[ \frac{1}{2}
\lambda_{ijk} {\hat L}^a_i {\hat L}^b_j {\hat E}_k^c
 - \mu_i {\hat L}^a_i \hat H_u^b \right] ~.
\label{lp}
\end{equation}

\n Moreover, if R-parity is not conserved then, in addition to the
usual RPC soft SUSY breaking terms, one must also add new
trilinear and bilinear soft terms corresponding to the RPV terms
of the superpotential, e.g., to the ones in (\ref{lp}). For our
purpose, the relevant ones to be added to the SUSY scalar
potential are the following soft breaking mass-like terms
\cite{GH,morebterms,davidson}:

\begin{equation}
V_{RPVBT} = (M_{LH}^2)_i \tilde L_i^a H_d^a -
\epsilon_{ab} b_i {\tilde L}^a_i H_u^b \label{bterm}~,
\end{equation}

\noindent where $\tilde L$ and $H_d$ are the scalar components of
$\hat L$ and $\hat H_d$, respectively.

\begin{figure}[htb]
\psfull
 \begin{center}
  \leavevmode
  \epsfig{file=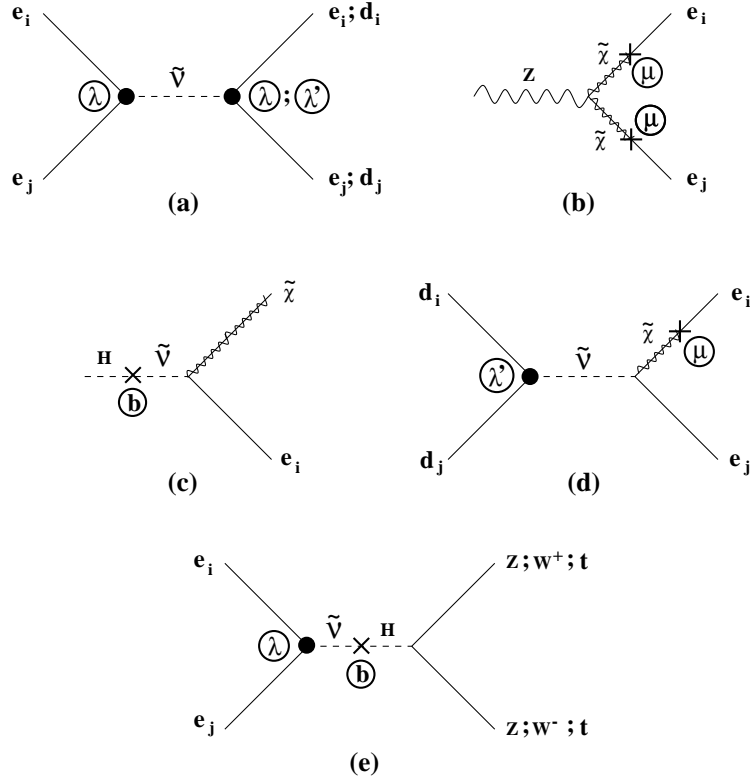,height=10cm,width=10cm,bbllx=0cm,bblly=2cm,bburx=23cm,bbury=25cm,angle=270}
 \end{center}
\caption{\emph{Examples of RPV signals according to the
combination of RPV couplings involved: (a) charged lepton or
down-quark pair production through a sneutrino resonance in
leptonic colliders $\propto \lambda \lambda$ or $\lambda
\lambda^\prime$, (b) leptonic flavor changing $Z$-decays, $Z \to
e_i \bar e_j$, $\propto \mu \mu$, (c) RPV scalar (e.g.,
Higgs) decay through sneutrino--Higgs mixing, e.g., $H \to \tilde\chi
e_i$ ($\tilde\chi$=chargino), $\propto b$, (d) rare Kaon flavor
changing leptonic decays, e.g., $K_L \to e_i \bar e_j$,
$\propto \lambda^\prime \mu$ and (e) the signal analyzed in this
paper: $s$-channel scalar resonance in massive gauge-boson and
top-quark pair production in leptonic colliders through
sneutrino--Higgs mixing, $\propto \lambda b$.}} \label{fig0}
\end{figure}

The presence of such lepton number violating interactions in the
SUSY Lagrangian drastically changes the phenomenology of SUSY. In
general the new phenomenological implications of a RPV SUSY
framework can be categorized according to the combinations of the RPV
couplings involved. For example, one of the most interesting
effects of the RPVTT in the superpotential is the possibility of
having an $s$-channel sneutrino resonance formation in fermion
pair production in leptonic colliders \cite{snures}. Such an
effect will be proportional either to the product $\lambda
\lambda$ (if a pair of leptons is produced) or to the product
$\lambda \lambda^\prime$ (if a pair of down-type quarks is
produced), where $\lambda$ and $\lambda^\prime$ are the trilinear 
RPV couplings in the superpotential (omitting subscripts),
see Fig.~\ref{fig0}(a). Among the interesting
phenomenological implication associated with the RPVBT in the
superpotential are tree-level neutrino masses
\cite{GH,basis,neutrinomass} and flavor changing $Z$-decays (see
e.g., Bisset {\it et al.}, in \cite{neutrinomass}); these effects
are, therefore, proportional to the products $\mu \mu$ (omitting subscripts), 
where $\mu$ 
are the bilinear RPV couplings in the superpotential, 
see Fig.~\ref{fig0}(b).

There are also new phenomena associated only with the soft RPVBT
of the scalar potential, i.e., with the $b$ terms in
(\ref{bterm}). Some examples for that are one-loop neutrino masses
\cite{GH,davidson,davidsonnew,GHnew} and new scalar decay channels
\cite{rpvbtdec}. These will be proportional to the soft bilinear
RPV coupling $b$ and arise as a consequence of mixings in the
scalar sector between the sleptons and the Higgs fields, see
Fig.~\ref{fig0}(c).

There are also RPV flavor changing phenomena in the SUSY fermionic
sector such as rare leptonic Kaon decays and radiative muon
decays, which are generated by the combined effect of the RPVTT
and RPVBT in the superpotential \cite{hepph0005262}. These type of
signals will be proportional to the products $\lambda \mu$ and/or
$\lambda^\prime \mu$, see Fig.~\ref{fig0}(d).

In a previous paper \cite{ourplb} we have suggested yet a new type
of signature of lepton number violation within SUSY, which is
proportional to the product of the RPVTT coupling $\lambda$ in the
superpotential and the soft breaking RPVBT coupling $b$ of the
SUSY scalar potential, i.e., $ \propto \lambda b$ - see
Fig.~\ref{fig0}(e). In particular, we have shown that one can have
an observable scalar--resonance enhancement in the cross-section
for producing a pair of massive gauge-bosons:

\begin{equation}
e^+e^- \to \Phi_{E,k} \to VV ~,~{\rm with}~~ V=W ~{\rm or}~ Z
\label{eevv}~,
\end{equation}

\n where $\Phi_{E,k}$, $k=1,2,3$, are admixtures of the RPC
CP-even neutral Higgs states and the sneutrino fields as
described below.\footnote{By an RPC state we
mean the state in the RPC limit of $b_3\to 0$.}

In this paper we wish to extend the analysis performed in
\cite{ourplb} and, in addition, to investigate the effect
of this type of RPV scalar resonances on top-quark pair
production:

\begin{equation}
e^+e^- \to \Phi_{E,k},\Phi_{O,\ell} \to t \bar t \label{eett}~.
\end{equation}

\n The new resonance signal in top-quark pair production
involves also $s$-channel
exchanges of the CP-odd scalar states (i.e., of $\Phi_{O,\ell}$,
$\ell=1,2,3$, which are admixtures of the RPC CP-odd neutral Higgs
and the CP-odd sneutrino fields), whereas in $VV$ production only
the CP-even admixtures ($\Phi_E$) contribute since the CP-odd
scalars do not couple to $VV$ at tree-level
\cite{higgshunters}.\footnote{We note that the simultaneous
presence of both CP-odd and CP-even scalar exchanges may give rise
to interesting tree-level CP-violating effects in $e^+e^- \to t
\bar t$. We do not discuss this possibility here.}

Such scalar resonances in top and massive gauge-boson pair
production can arise with measurable consequences when the
incoming $e^+e^-$ beams couple to the sneutrino component in the
new physical states, $\Phi_{E}$ or $\Phi_O$, with a coupling
$\lambda \gg m_e/M_W$ (that is the typical Higgs coupling to the 
incoming beams) in (\ref{lp}), while the $VV$ and
the $t \bar t$ final states couple to the Higgs components either in
$\Phi_{E}$ or in $\Phi_O$. Then, a non-vanishing scalar
resonance in $e^+ e^- \to VV$ and $e^+ e^- \to t \bar t$ can be 
attributed to the Higgs--sneutrino mixing phenomena, and can serve
as an exclusive probe of the soft breaking RPVBT in the SUSY
scalar potential (i.e., of $b$). Indeed, the resonance effects are
essentially proportional to the product $\lambda b$ which will
vanish as $b \to 0$.

This differs from the situation of down-quark pair
production via $e^+e^- \to \Phi_{E,k},\Phi_{O,\ell} \to d \bar d$
($d=d,~s,~b$) where the RPV scalar resonance may occur in two
ways: ($a$) when the sneutrino component in $\Phi_{E}$ or in
$\Phi_O$ couples to both the $e^+e^-$ and $d \bar d$ through its
trilinear RPV couplings $\lambda$ and $\lambda^\prime$,
respectively (see e.g., \cite{RPVeebb} for a discussion of such a
sneutrino resonance effect in $e^+e^- \to b \bar b$); ($b$) when the
sneutrino component couples to $e^+e^-$ then mixes with the Higgs
state which couples to the $d \bar d$ final state. It is clear
that in down quark pair production such a scalar resonance
will be dominated by the purely RPVTT effect of type ($a$) above, as
long as $\lambda^\prime$ is larger than the small Yukawa coupling
of the neutral Higgs to down-quarks. In particular, an observable
scalar resonance enhancement in $d \bar d$ production is
possible even when $b \to 0$ and only the RPVTT contributes. On
the other hand, in the $t \bar t$ production channel, the purely
RPVTT effect of type (1) above is absent at tree-level (the $t
\bar t \tilde\nu$ coupling is forbidden by gauge-invariance).
Similarly, there is no tree-level $VV \tilde\nu$ coupling when the
soft breaking RPVBT vanishes, i.e., when $b \to 0$.

Thus, in contrast with down-quark and charged-lepton pair
production in which a measurable scalar resonance enhancement is
expected to emanate from the purely trilinear RPV couplings, in
the channels considered here, the scalar resonance enhancement
will arise only if the sneutrino and the Higgs states mix via the
soft RPVBT. We therefore stress again that the scalar resonance
formations in $t \bar t$ and $VV$ production are fundamentally
different from previously suggested sneutrino resonances in
leptonic colliders within RPV SUSY (e.g., in fermion pair
production \cite{snures}), since they are driven by RPV parameters
from the soft breaking scalar sector and not purely by Yukawa-like
RPV couplings from the superpotential. It should also be noted
that the mechanism of a scalar resonance via sneutrino--Higgs
mixing that we are considering here 
will not be efficient for light up-quark (i.e., u and
c-quarks) pair production due to the smallness of the
corresponding Higgs-$u$-$\bar u$ and the Higgs-$c$-$\bar c$ Yukawa
couplings.

The paper is organized as follows: in Section II we define our low
energy RPV SUSY framework and assumptions, we present the CP-even
and CP-odd scalar mass matrices in the presence of the soft RPVBT
and discuss their behavior in some limiting cases. In section
III we derive the relevant Feynman rules for the new scalar
mass-eigenstates and we calculate the cross-sections for the
$s$-channel scalar exchanges in $e^+e^- \to ZZ,~WW$ and $e^+e^-
\to t \bar t$. In section IV we present our numerical results for
the expected resonance signals in $e^+e^- \to VV$, scanning the
relevant SUSY parameter space at LEP2 and future linear colliders. 
In section V we investigate the
sensitivity of future linear colliders to such
scalar resonance signals in $e^+e^- \to t \bar t$. In section
VI we present our conclusions.

\section{Notations, assumptions and features of the sneutrino--Higgs mixing phenomena}

In what follows, for simplicity we will assume $b_i \neq
0$ only when $i=3$ in (\ref{bterm}), thus, considering only the mixing between the
stau ($\tilde L_3$) and the Higgs scalar fields ($H_d$ and $H_u$).
The consequences of $b_1 \neq 0$ and/or $b_2 \neq 0$ is to
introduce additional mixings among sleptons of different
generations and mixings between the selectron and/or smuon with
the Higgs fields which are not crucial for the main outcome of
this paper.

The bilinear soft term $b_3$ leads in general to a non-vanishing
VEV of the tau-sneutrino, $\langle \tilde\nu_3 \rangle = v_3 \neq
0$. However, since lepton number is not a conserved quantum number
in this scenario, the $\hat H_d$ and $\hat
L_3$ superfields lose their identity and can be rotated to a
particular basis ($\hat H_d^\prime,\hat L_3^\prime$) in which
either $\mu_3$ or $v_3$ are set to zero
\cite{GH,basis,davidson,davidsonnew,GHnew}. In what follows, we
find it convenient to choose the ``no VEV'' basis, $v_3=0$, which
simplifies our analysis below.

Furthermore, our key assumption will be that the lepton number
violating couplings in the SUSY Lagrangian are small compared to
the corresponding RPC ones wherever they appear. More
specifically, the RPV parameters, $\lambda$ and $b_3$, will be
limited such that $|\lambda_{ijk}| \le 0.1$ and $b_3/b_0 \le 0.1$.
It is worth noting that the minimization of the scalar potential
yields (in the $v_3=0$ basis) \cite{GH}: $b_3=(M^2_{\tilde L H} +
\mu_3 \mu_0) / t_\beta$, where $\tan\beta \equiv v_u/v_d$. Thus,
in the general case, $b_3$ needs not vanish even if $\mu_3$ is
vanishingly small, as may be suggested by low energy flavor
changing processes (see e.g., \cite{hepph0005262}) and flavor
changing $Z$-decays (see e.g., Bisset {\it et al.}, in
\cite{neutrinomass}). In particular, if $\mu_3 \to 0$ (so that
$M_{\tilde L H}^2 \gg \mu_3 \mu_0$) then $b_3 \sim M_{\tilde L
H}^2 /t_\beta$, in which case RPV in the scalar potential
decouples from the RPV in the superpotential (i.e., $b_3$ is
independent of $\mu_3$). In such a case, small lepton number
violation in the scalar potential should be realized by requiring
only that $b_3 \ll b_0$.\footnote{Note that the laboratory limit
on the $\tau$-neutrino mass allows $b_3/b_0 \sim {\cal O}(1)$
\cite{davidson}} In this paper we will adopt this approach which
treats RPV in the scalar potential independently from RPV in the
superpotential.

Let us define the SU(2) components of the up and down neutral
Higgs and tau-sneutrino scalar fields:\footnote{We always use the
superscript $0$ to denote what would be the scalar states in the
RPC limit $b_3 \to 0$.}

\begin{eqnarray}
H^0_{u} &\equiv& (\xi_{u}^0 + v_{u} + i \phi_{u}^0)/\sqrt{2} ~,\nonumber \\
H^0_{d} &\equiv& (\xi_{d}^0 + v_{d} + i \phi_{d}^0)/\sqrt{2}
\label{components}~, \\
{\tilde\nu_\tau} &\equiv& (\tilde\nu_+^0 + v_3 + i
\tilde\nu_-^0)/\sqrt{2} \nonumber ~,
\end{eqnarray}

\n where, as stated above, in the following discussion we always
set $v_3=0$.

The CP-even and CP-odd $3\times 3$ symmetric
scalar squared-mass matrices $M_E^2$ and $M_O^2$ respectively,
are then obtained
through the quadratic part of the scalar potential:

\begin{eqnarray}
\frac{1}{2} \left(\Phi^0_{E,O} \right)^T
M_{E,O}^2 \Phi^0_{E,O} ~,
\end{eqnarray}

\n where

\begin{eqnarray}
\Phi^0_E =\left( \xi_d^0,\xi_u^0,\tilde\nu_+^0 \right) ~~{\rm
and}~~ \Phi^0_O = \left( \phi_d^0,\phi_u^0,\tilde\nu_-^0 \right)~,
\end{eqnarray}

\n are the SU(2) weak states. The new CP-even and CP-odd scalar
mass-eigenstates (i.e., the physical states) are derived by
diagonalizing $M_{E,O}^2$. Let us denote the physical states by:

\begin{eqnarray}
\Phi_E =\left( H,h,\tilde\nu_+ \right) ~~,~~
\Phi_O = \left( A,G,\tilde\nu_- \right)~,
\end{eqnarray}

\n such that for small RPV in the SUSY Lagrangian, $H,~h$ and
$\tilde\nu_+$ are the states dominated by the CP-even RPC states
$H^0,~h^0$ and $\tilde\nu_+^0$, respectively, and $A,~\tilde\nu_-$
are the states dominated by the CP-odd RPC states
$A^0,~\tilde\nu_-^0$, respectively. Also, $G$ is the
Goldstone boson that is absorbed by the $Z$-boson and, therefore,
is the state with a zero eigenvalue in $M_O^2$.

Thus, the physical states ($\Phi_{E,O}$) are related to the weak
eigenstates ($\Phi_{E,O}^0$) via $\Phi^0_{E,O} = S_{E,O}
\Phi_{E,O}$, where $S_{E,O}$ are the rotation matrices that
diagonalize $M_{E,O}^2$:

\begin{eqnarray}
S^T_{E} M_{E}^2 S_{E}= \pmatrix{ m_H^2& 0 & 0 \cr 0 & m_h^2 & 0
\cr 0 & 0 & (m_{s \nu}^+)^2 } \label{semese}~,
\end{eqnarray}

\begin{eqnarray}
S^T_{O} M_{O}^2 S_{O}= \pmatrix{ m_A^2& 0 & 0 \cr 0 & 0 & 0 \cr 0
& 0 & (m_{s \nu}^-)^2 } \label{somoso}~.
\end{eqnarray}

\n As in (\ref{semese}) and (\ref{somoso}) and 
throughout the rest of the paper, we will denote by
$m_H,m_h,m_{s\nu}^+$ and $m_A,m_{s\nu}^-$ the masses of the
CP-even and CP-odd physical states (when $b_3 \neq 0$),
respectively. Similarly, adding the superscript $0$,
$m_H^0$,$m_h^0$,$m_{s \nu}^0$,$m_A^0$ will denote the corresponding masses
in the RPC limit ($b_3 \to 0$). Note that in the RPC limit the
CP-even and CP-odd tau-sneutrino states, $\tilde\nu_+^0$ and
$\tilde\nu_-^0$, do not mix, and are, therefore, degenerate with a
common mass $m_{\tilde\nu_+^0}=m_{\tilde\nu_-^0} \equiv m_{s
\nu}^0$.

In the RPC limit the Higgs and sneutrino sectors decouple. That
is, $M_E^2$ and $M_O^2$ consist of the usual $2\times 2$ upper
left blocks corresponding to the two CP-even and CP-odd Higgs
states, respectively (which can be described at tree-level by only
two parameters \cite{gunion,djuadi}) plus one sneutrino entry. 
The two Higgs parameters are 
conventionally chosen to be
$m_A^0$ and $t_\beta \equiv \tan\beta$, where $m_A^0$ is related
at tree-level to the soft bilinear mass term $b_0$ via
$b_0=(m_A^0)^2 c_\beta s_\beta$ ($s_\beta \equiv \sin\beta$ and
$c_\beta \equiv \cos\beta$). Therefore, if the scalar potential conserves
R-parity, then $M_E^2$ and $M_O^2$ can be written (at
tree-level) as \cite{GH,gunion,mhcorrections}:

\begin{eqnarray}
M^2_E(RPC) = \pmatrix{(m_A^0)^2 s_\beta^2 + m_Z^2 c_\beta^2 &
-\left[ (m_A^0)^2 + m_Z^2 \right] s_\beta c_\beta & 0 \cr -\left[
(m_A^0)^2 + m_Z^2 \right] s_\beta c_\beta & (m_A^0)^2 c_\beta^2 +
m_Z^2 s_\beta^2 & 0 \cr 0 & 0 & (m_{s \nu}^0)^2 }~,
\end{eqnarray}

\begin{eqnarray}
M^2_O(RPC) = \pmatrix{(m_A^0)^2 c_\beta^2 & (m_A^0)^2 c_\beta
s_\beta & 0 \cr (m_A^0)^2 c_\beta s_\beta & (m_A^0)^2 s_\beta^2 &
0 \cr 0 & 0 & (m_{s \nu}^0)^2 } ~.
\end{eqnarray}

\n However, if lepton number is violated in the scalar potential
through $b_3 \neq 0$, then $M_E^2$ and $M_O^2$ acquire non-zero
off-diagonal $\xi^0_{d,u}-\tilde\nu_+^0$ and
$\phi^0_{d,u}-\tilde\nu_-^0$ mixing entries, respectively (which
are $\propto b_3$), while in the "no VEV" basis the pure $2 \times
2$ Higgs blocks remain unchanged \cite{GH}. In this case, $M_E^2$ and
$M_O^2$ are given by:

\begin{eqnarray}
M^2_E = \pmatrix{(m_A^0)^2 s_\beta^2 + m_Z^2 c_\beta^2 +
\delta_{dd} & -\left[ (m_A^0)^2 + m_Z^2 \right] s_\beta c_\beta +
\delta_{du} & b_3 t_\beta \cr -\left[ (m_A^0)^2 + m_Z^2 \right]
s_\beta c_\beta +\delta_{du} & (m_A^0)^2 c_\beta^2 + m_Z^2
s_\beta^2 + \delta_{uu} & -b_3 \cr b_3 t_\beta & -b_3 & (m_{s
\nu}^0)^2 } \label{me2}~,
\end{eqnarray}

\begin{eqnarray}
M^2_O = \pmatrix{(m_A^0)^2 c_\beta^2 & (m_A^0)^2 c_\beta s_\beta &
b_3 t_\beta \cr (m_A^0)^2 c_\beta s_\beta & (m_A^0)^2 s_\beta^2 &
 b_3 \cr
 b_3 t_\beta & b_3 & (m_{s \nu}^0)^2 } \label{mo2}~.
\end{eqnarray}

\n In (\ref{me2}) we have symbolically added to the tree-level
CP-even mass matrix $M_E^2$ the quantities
$\delta_{dd},~\delta_{du}$ and $\delta_{uu}$ which are the 1-loop
corrections to the CP-even pure Higgs block, i.e., the
$(\xi_d^0,\xi_u^0)$ block. Indeed, in our numerical analysis
we include in $\delta_{dd},~\delta_{du}$ and $\delta_{uu}$
the dominant 1-loop corrections coming from the $t - \tilde t$
sector. Below we use the present LEP2 lower bound on
$m_h$ in order to place limits on our RPV parameter space.
The higher order corrections to the tree-level
CP-even Higgs sector are, therefore, essential in evaluating the true
theoretical value of the light Higgs mass, $m_h$, since they can
generate a significant deviation (up to $50\%$) to the tree-level
value of $m_h$.

Let us further note the following:
\begin{itemize}

\item Above we have "traded" the RPC soft
breaking bilinear mass term $b_0$ for the ``bare'' pseudo-scalar
Higgs mass $m_A^0$ (i.e., what would have been the physical mass
of the pseudo-scalar Higgs state if R-parity were conserved), by using
the RPC tree-level relation $b_0=(m_A^0)^2 s_\beta c_\beta$ which,
for $t_\beta^2 \gg 1$, implies $(m_A^0)^2 \sim b_0 t_\beta$. As
will be shown below, for small RPV ($b_3/b_0 \ll 1$), $A^0$ and
the new pseudo-scalar mass-eigenstate $A$ are almost degenerate
(i.e., $m_A \sim m_A^0$) if no accidental mass degeneracy occurs.
This clearly follows from our definition of quantities which are
denoted with the superscript $0$. Hence, with the assumption of
small RPV (${\rm RPV/RPC} \ll 1$) $m_A$ also scales as $(m_A)^2
\sim b_0 t_\beta$, for $t_\beta^2 \gg 1$. Without loss of
generality we then set:

\begin{equation}
b_3 \equiv \varepsilon (m_A^0)^2 /t_\beta \sim \varepsilon m_A^2 /t_\beta
\label{epsdef}~,
\end{equation}

\n such that the small lepton number violation in the scalar sector is
parameterized by the dimensionless quantity $\varepsilon \sim
b_3/b_0$. Then $\varepsilon \ll 1$ corresponds to $b_3 \ll b_0$.

\item As can be seen from (\ref{me2}) and (\ref{mo2}),
as a result of $b_3\neq 0$, the usual CP-even RPC Higgs states
$H^0$ and $h^0$ ($m_{H^0}
 > m_{h^0}$) will acquire a small
$\tilde\nu_+^0$ component and vice
versa due to
the non-zero $(M_E^2)_{13,23,31,32}$ elements.
Similarly, due to $(M_O^2)_{13,23,31,32} \neq 0$,
the RPC pseudo-scalar state $A^0$ will acquire a
small $\tilde\nu_-^0$ component and vice versa.

\item Setting $b_3 \equiv \varepsilon (m_A^0)^2 /t_\beta$,
our relevant low-energy SUSY parameter space is fully determined
at tree-level by the four parameters $m_A^0$, $m_{s
\nu}^0$, $t_\beta$ and $\varepsilon$. In particular, these four
parameters completely fix $M_{E}^2$ and $M_O^2$ at tree-level from
which the CP-even and CP-odd rotation matrices $S_E$ and $S_O$ as
well as the tree-level physical masses $m_{\Phi_{E,k}}$ and
$m_{\Phi_{O,k}}$ are derived by the diagonalization procedure.

\end{itemize}

Assuming that $\varepsilon$ is a small parameter, we solve the
eigenvalues equations for $M_E^2$ and $M_O^2$ perturbatively up to the
second order in $\varepsilon$. We can thus write the new
(physical) mass-squared eigenvalues, $m_{\Phi_{E,k}}^2$ and
$m_{\Phi_{O,k}}^2$, in terms of the corresponding eigenvalues in
the RPC limit, $m_{\Phi_{E,k}^0}^2$ and $m_{\Phi_{O,k}^0}^2$, as
follows:

\begin{eqnarray}
m_{\Phi_{E,k}}^2 &=& m_{\Phi_{E,k}^0}^2 \left( 1 +
 \varepsilon \delta_{E,k}^{(1)} +  \varepsilon^2 \delta_{E,k}^{(2)}  +
{\cal O}( \varepsilon^3) \right) ~, \\
m_{\Phi_{O,k}}^2 &=& m_{\Phi_{O,k}^0}^2 \left( 1 +
 \varepsilon \delta_{O,k}^{(1)} +  \varepsilon^2 \delta_{O,k}^{(2)}  +
{\cal O}( \varepsilon^3) \right) ~.
\end{eqnarray}

We then find that $\delta_{E,k}^{(1)}=\delta_{O,k}^{(1)}=0$ and

\begin{eqnarray}
\delta_{E,k}^{(2)}
&\propto& \left(\frac{ m_A^0 }{ m_{\Phi_{E,k}^0} }\right)^2 \times
\frac{(m_A^0)^2}{ m_{\Phi_{E,k}^0}^2 - m_{\Phi_{E,\ell}^0}^2 } ~;~
k \neq \ell \label{shift1}~,\\
\delta_{O,k}^{(2)}
&\propto& \left(\frac{ m_A^0 }{ m_{\Phi_{O,k}^0} }\right)^2 \times
\frac{(m_A^0)^2}{ m_{\Phi_{O,k}^0}^2 - m_{\Phi_{O,\ell}^0}^2 } ~;~
k \neq \ell \label{shift2}~.
\end{eqnarray}

\n Thus, with
$\varepsilon \ll 1$ (implying $b_3 \ll b_0$ and also $b_3 \ll (m_A^0)^2$),
the mass shifts induced by the RPVBT in (\ref{bterm})
are, at leading order, $\propto \varepsilon^2$
(see also \cite{GH}).
It is evident from (\ref{shift1}) and (\ref{shift2}) that,
as long as there are
no accidental mass degeneracies among
scalar states of the same CP property,
these mass shifts will remain small for $\varepsilon \ll 1$.
Hence, although we are using
``bare'' masses (i.e., the scalar masses in the RPC
limit) as inputs, it should be kept in mind that the physical masses are
slightly shifted if there are no mass degeneracies as illustrated next.

\begin{figure}[htb]
\psfull
 \begin{center}
  \leavevmode
  \epsfig{file=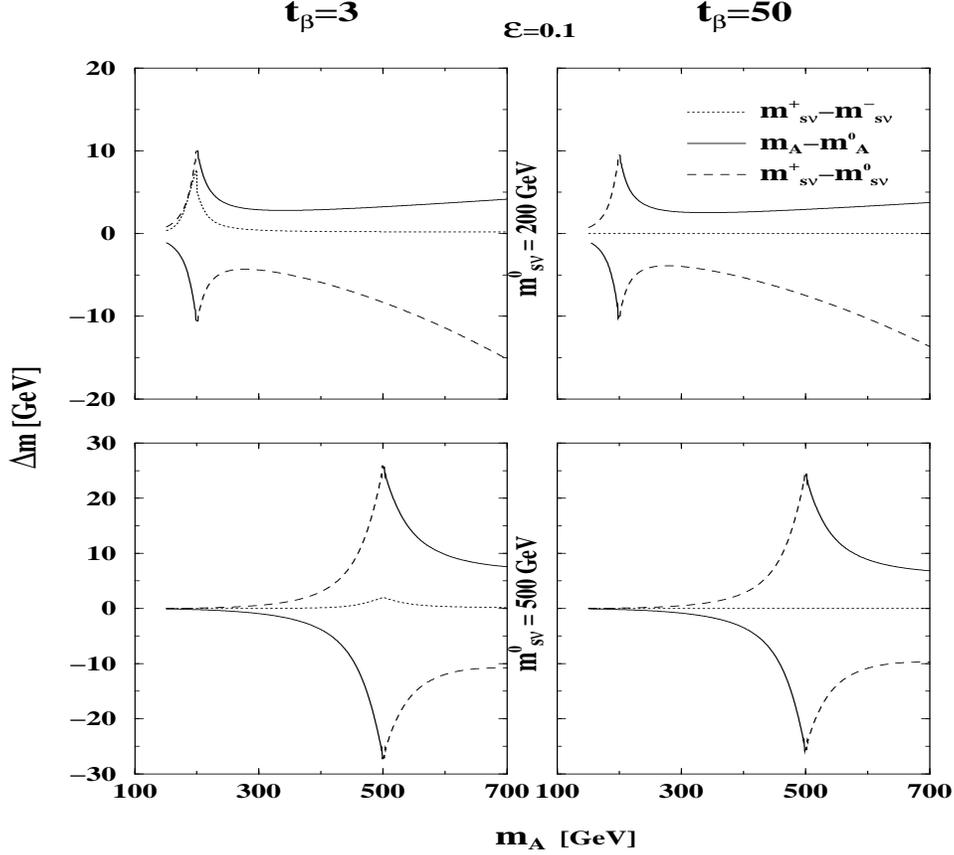,height=11cm,width=13cm,bbllx=0cm,bblly=1cm,bburx=20cm,bbury=26cm,angle=0}
 \end{center}
\caption{\emph{The mass shifts $\Delta m_A \equiv m_A - m_A^0$,
$\Delta m_{s \nu}^- \equiv m_{s \nu}^- - m_{s \nu}^0$ and the
sneutrino mass splitting $\Delta m_{s \nu}^\pm \equiv m_{s \nu}^+
- m_{s \nu}^-$ induced by $\varepsilon = 0.1$, as a function of
the RPC pseudo-scalar mass $m_A^0$, for two values of the RPC
sneutrino mass $m_{s \nu}^0=200$ and $500$ GeV and for $t_\beta=3$
(left figures) or $t_\beta=50$ (right figures). See also text.}}
\label{fig1}
\end{figure}

In Fig.~\ref{fig1} we plot the mass shifts in the pseudo-scalar
states: $\Delta m_A \equiv m_A - m_A^0$, $\Delta m_{s \nu}^-
\equiv m_{s \nu}^- - m_{s \nu}^0$ and the mass splitting between
the two CP-even and CP-odd sneutrino states, $\Delta m_{s \nu}^\pm
\equiv m_{s \nu}^+ - m_{s \nu}^-$, due to $\varepsilon \neq 0$.
This is shown as a function of $m_A^0$, for two representative
values of the "bare" sneutrino mass $m_{s \nu}^0=200,~500$ GeV,
for $\varepsilon=0.1$ and $\tan\beta=3$ or 50. As expected from
(\ref{shift1}) and (\ref{shift2}), $\Delta m_A$ and $\Delta m_{s
\nu}^-$ increase with $m_A^0$ and are inversely proportional to
$(m_{s \nu}^0 - m_A^0)$ (they change sign at the turning point
$m_{s \nu}^0 = m_A^0$). Also, for $\varepsilon=0.1$ and in the
ranges of $m_A^0$ and $m_{s \nu}^0$ considered, $\Delta m_A/m_A^0$
and $\Delta m_{s \nu}^-/m_{s \nu}^0$ are typically at the level of
a few percent for both a low and a high $\tan\beta$ scenario
(even for almost degenerate $m_A^0$ and $m_{s \nu}^0$).

As for the sneutrino mass splitting $\Delta m_{s \nu}^\pm$, we see
that a non-vanishing $\tilde\nu_+ - \tilde\nu_-$ mixing can occur only for
a low $\tan\beta$ scenario,
and when $m_A^0$ and $m_{s \nu}^0$ are sufficiently close.
It should be noted that
this sneutrino mixing phenomena can give rise to interesting
lepton number violating effects such as radiative neutrino masses
\cite{davidsonnew,GHnew,snumix} and tree-level
CP-violation in fermion pair production
\cite{ourtautaupaper}.

The behavior of the mass shifts in the CP-even sector,
$\Delta m_{s \nu}^+ \equiv m_{s \nu}^+ - m_{s \nu}^0$ and
$\Delta m_H \equiv m_H - m_H^0$, although
not shown, is easily traced. Indeed, $\Delta m_{s \nu}^+$ may be
inferred from the combination of $\Delta m_{s \nu}^-$ and
$\Delta m_{s \nu}^\pm$ shown in Fig.~\ref{fig1}:
for very large $\tan\beta$ values,
$\Delta m_{s \nu}^+ \sim \Delta m_{s \nu}^-$
whereas for smaller $\tan\beta$ values,
$\Delta m_{s \nu}^+$ approaches $\Delta m_{s \nu}^-$ as
$m_A^0$ is increased.
In the case of the heavier CP-even Higgs, we find that
$\Delta m_H \sim \Delta m_A$
in the decoupling limit where $(m_A^0)^2 \gg M_Z^2$, since then
$m_H^0 \sim m_A^0$.

As mentioned above, due to the existing LEP2 lower bounds
on the mass of the light CP-even Higgs, $h$, the mass shift
$\Delta m_h = m_h - m_h^0$ caused by the RPVBT operator in
(\ref{bterm}), is crucial, since it determines the
allowed range of our free parameter space $\left\{
\varepsilon,~m_A^0,m_{s \nu}^0,~\tan\beta \right\}$ on which
$\Delta m_h$ depends. Moreover, the fact that the mass shifts
in the CP-even sector are proportional to the sign of
($m_{\Phi_{E,k}^0} - m_{\Phi_{E,\ell}^0}$) has important
consequences on the light CP-even Higgs particle. In particular,
we find that if
$m_A^0,m_{s \nu}^0 > m_{h^0}$ (as always chosen below), then $m_h$
tends to decrease with $\varepsilon$. We can thus use the present
LEP2 limit on $m_h$ to deduce the allowed range in e.g., the
$\varepsilon - m_{s \nu}^0$ plane, for a given $m_A^0$ (e.g., for
$m_A \gsim 200$ GeV the present LEP2 bound is roughly $m_h \gsim
110$ GeV, irrespective of $t_\beta$ and in the maximal mixing
scenario with a typical SUSY scale/squark mass of 1 TeV
\cite{lep2mh})\footnote{Since $b_3 \ne 0$ the $hZZ$ coupling is
smaller than its value in the RPC case leading to a smaller
$e^+e^- \to Zh$ production rate. The limits on $m_h$ given in
\cite{lep2mh} are therefore slightly weaker in the RPV case (see
also \cite{davidson}).}. Therefore, as mentioned earlier, we
include the dominant higher order corrections (coming from the $t
- \tilde t$ sector) to the ($\xi_d^0,\xi_u^0$) block in $M_E^2$, which
are denoted by $\delta_{uu},\delta_{dd}$ and $\delta_{du}$
in (\ref{me2}). To do that, we use the approximated formulae given in
\cite{mhcorrections} with the maximal mixing scenario (as defined
in \cite{mhcorrections}), and set the typical squark mass at
$m_{\tilde q} \sim 1$ TeV.

\begin{figure}[htb]
\psfull
 \begin{center}
  \leavevmode
  \epsfig{file=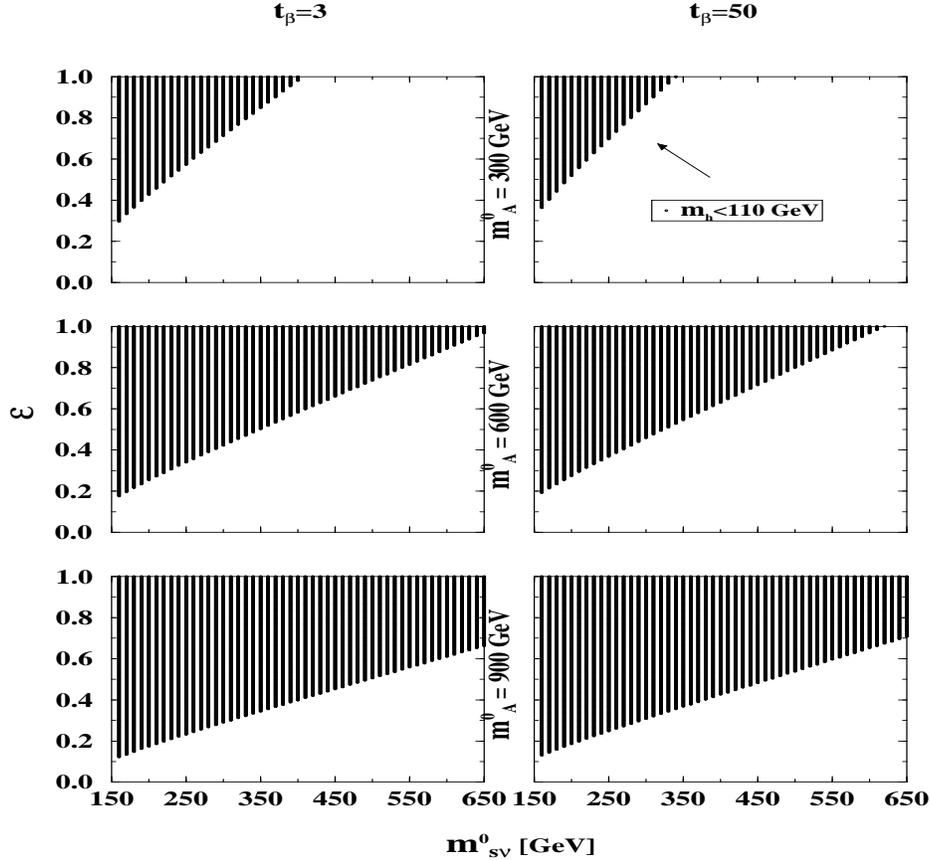,height=11cm,width=13cm,bbllx=0cm,bblly=1cm,bburx=20cm,bbury=26cm,angle=0}
 \end{center}
\caption{\emph{The shaded areas in the $\varepsilon - m_{s \nu}^0$ plane
[$\varepsilon \equiv b_3 t_\beta/ (m_A^0)^2$]
are excluded by the recent LEP2 limit on the light Higgs
mass $m_h \gsim 110$ GeV. These
excluded regions are given for $m_A^0=300,~600$ or $900$ GeV, for
$t_\beta=3$ (left figures) or $t_\beta=50$ (right figures) and are
 independent of the trilinear RPV coupling $\lambda_{131}$.}}
\label{fig2}
\end{figure}

In Fig.~\ref{fig2} we show the excluded region in the
$\varepsilon - m_{s \nu}^0$ plane (the shaded area)
from the recent LEP2 limit of $m_h \gsim 110$ GeV which holds for
the parameters set
$\tan\beta=3$ or 50
and $m_A^0=300,~600$ or $600$ GeV assumed in Fig.~\ref{fig2}.

In what follows, in addition to our assumption of small RPV,
we will focus on the case of a heavy Higgs
spectrum (sometimes referred to as the decoupling limit) with
$(m_A^0)^2 \gg M_Z^2$. As noted above,
in this case there is a near
mass degeneracy among the heavy CP-even Higgs and the pseudoscalar
Higgs, $m_H^0 \sim m_A^0$ in the RPC limit. As shown in Fig.~\ref{fig1}, as
long as $\varepsilon \ll 1$, this near mass
degeneracy will also hold among the corresponding new physical
states, i.e., $m_H \sim m_A$.

Therefore, we wish to further investigate the behavior of the
mixing matrices $S_E$ and $S_O$ under the three assumptions:
($a$) $\varepsilon \ll 1$ (small RPV in the scalar sector), ($b$)
$(m_A^0)^2 \gg M_Z^2$ (heavy Higgs spectrum) and ($c$) $t_\beta^2
\gg 1$. In the small RPV scenario the elements of the CP-even and
CP-odd mixing matrices, $S_E$ and $S_O$, approach their RPC values:

\begin{eqnarray}
S_E ~ \stackrel{\varepsilon \to 0} {\longrightarrow} ~ S_E^0 =
\pmatrix{ c_\alpha & - s_\alpha & 0 \cr s_\alpha & c_\alpha & 0
\cr 0 & 0 & 1 } \label{seeps}~,
\end{eqnarray}

\begin{eqnarray}
S_O ~ \stackrel{\varepsilon \to 0} {\longrightarrow} ~ S_O^0 =
\pmatrix{ s_\beta & - c_\beta & 0 \cr c_\beta & s_\beta & 0 \cr 0
& 0 & 1 } \label{soeps}~,
\end{eqnarray}

\noindent where $S_E^0$ and $S_O^0$ denote the corresponding
CP-even and CP-odd scalar mixing matrices when R-parity is
conserved. Also, $s_\alpha(c_\alpha) \equiv \sin\alpha(\cos\alpha)$ and
$\alpha$ is the usual mixing angle of the RPC CP-even neutral
Higgs sector defined through \cite{rosiek,gunion}:\footnote{Note
that the 1-loop corrections to the pure Higgs block in the CP-even
sector slightly shift the mixing angle, i.e., $\alpha \to
\alpha^\prime$. While these 1-loop effects are always included in
our numerical analysis, for the purpose of understanding the
qualitative features of the sneutrino--Higgs mixing phenomena it
suffices to consider the tree-level values of $M_E^2$. The small
shift in $\alpha$ generates a significant effect only for the
light CP-even Higgs $h$.}

\begin{eqnarray}
\tan 2 \alpha= \tan 2 \beta \frac{(m_A^0)^2 + M_Z^2}{(m_A^0)^2 -
M_Z^2} ~.
\end{eqnarray}

\n Consider now the matrices $S_E^0$ and $S_O^0$ in the decoupling
limit $(m_A^0)^2 \gg M_Z^2$. Since in this
limit the heavier CP-even Higgs RPC state, $H^0$, and the CP-odd
RPC state, $A^0$ are almost degenerate,
both $H^0$ and $A^0$ are considerably heavier than the
$Z$-boson and, therefore, also heavier than the lighter CP-even
Higgs state $h^0$. In particular, when $(m_A^0)^2 \gg M_Z^2$ one
obtains \cite{djuadi}:

\begin{eqnarray}
\cos\alpha ~ \stackrel{(m_A^0)^2 \gg M_Z^2} {\longrightarrow} ~
\sin\beta ~~~,~~~ \sin\alpha ~ \stackrel{(m_A^0)^2 \gg M_Z^2}
{\longrightarrow} ~ - \cos\beta ~,
\end{eqnarray}

\n which in turn yields in the CP-even scalar sector:

\begin{eqnarray}
S_E^0 ~ \stackrel{(m_A^0)^2 \gg M_Z^2} {\longrightarrow} ~
\pmatrix{ s_\beta & c_\beta & 0 \cr - c_\beta & s_\beta & 0 \cr 0
& 0 & 1 } \label{sedec}~.
\end{eqnarray}

If in addition $t_\beta^2 \gg 1$, then it follows from 
(\ref{sedec}) that:

\begin{eqnarray}
&&\xi^0_d ~\stackrel{\varepsilon \to 0} {\longrightarrow}~
c_\alpha H - s_\alpha h  \stackrel{(m_A^0)^2 \gg
M_Z^2}{\longrightarrow}~ s_\beta H + c_\beta h \stackrel{t_\beta^2
\gg 1}{\longrightarrow}~ H \label{rel1} ~,\\
&&\xi^0_u ~\stackrel{\varepsilon \to 0} {\longrightarrow}~
s_\alpha H + c_\alpha h  \stackrel{(m_A^0)^2 \gg
M_Z^2}{\longrightarrow}~ -c_\beta H + s_\beta h
\stackrel{t_\beta^2 \gg 1}{\longrightarrow}~ h \label{rel2}~,
\end{eqnarray}

\n where now $H$ and $h$ are the physical states so that
$H$($h$) is mostly composed out of the $\xi_d^0$($\xi_u^0$) weak
state.
Similarly, under $\varepsilon \ll 1$ and $t_\beta^2 \gg 1$,
(\ref{soeps}) will yield:

\begin{eqnarray}
&&\phi^0_d ~\stackrel{\varepsilon \to 0} {\longrightarrow}~
s_\beta A - c_\beta G  \stackrel{t_\beta^2 \gg 1}{\longrightarrow}~
A \label{relo1} ~,\\
&&\phi^0_u ~\stackrel{\varepsilon \to 0} {\longrightarrow}~
c_\beta A + s_\beta G  \stackrel{t_\beta^2 \gg 1}{\longrightarrow}~ G
 \label{relo2}~,
\end{eqnarray}

\n so that $A$ is dominated by the $\phi_d^0$ weak state.

For a summary of the notation introduced in this section for the various 
scalar states see Table \ref{tab1}. 

\begin{table}[htb]
\begin{tabular}{||c|c||}
Symbol & Particle/State it represents \\ \hline \hline
$H_u^0$ & Neutral field of the initial up-type Higgs SU(2) doublet \\ 
\hline
$H_d^0$ & Neutral field of the initial down-type Higgs SU(2) doublet \\ 
\hline
$\tilde\nu_\tau$ & Neutral field of the initial 3rd generation SU(2) 
slepton doublet \\ 
\hline
$\xi_u^0$ & CP-even component of $H_u^0$ \\ 
\hline
$\phi_u^0$ & CP-odd component of $H_u^0$ \\ 
\hline
$\xi_d^0$ & CP-even component of $H_d^0$ \\ 
\hline
$\phi_d^0$ & CP-odd component of $H_d^0$ \\ 
\hline
$\tilde\nu_+^0$ & CP-even component of $\tilde\nu_\tau$,  physical state when $b_3 = 0$ \\ 
\hline
$\tilde\nu_-^0$ & CP-odd component of $\tilde\nu_\tau$, physical state when $b_3 = 0$  \\ 
\hline
$\Phi_E^0$ & CP-even weak states, $\Phi_E^0 = \xi_u^0,~\xi_d^0$ or 
$\tilde\nu_+^0$ \\ 
\hline
$\Phi_O^0$ & CP-odd weak states, $\Phi_O^0 = \phi_u^0,~\phi_d^0$ or 
$\tilde\nu_-^0$ \\ 
\hline
$h^0$ & CP-even light Higgs, physical state when $b_3 = 0$ \\ \hline
$H^0$ & CP-even heavy Higgs, physical state when $b_3 = 0$  \\ \hline
$A^0$ & CP-odd Higgs, physical state when $b_3 = 0$  \\ \hline
$h$ & CP-even light Higgs, physical state for any (value of) $b_3$ \\ \hline
$H$ & CP-even heavy Higgs, physical state for any (value of) $b_3$ \\ \hline
$A$ & CP-odd  Higgs, physical state  for any (value of) $b_3$ \\ \hline
$G$ & CP-odd  Goldstone boson \\ \hline
$\tilde\nu_+$ & CP-even $\tau$-sneutrino, physical state for any (value of) $b_3$ \\ \hline  
$\tilde\nu_-$ & CP-odd $\tau$-sneutrino, physical state for any (value of) $b_3$ \\ \hline  
$\Phi_E$ & CP-even physical states for any (value of) $b_3$, $\Phi_E = H,~h$ or $\tilde\nu_+$ \\ \hline
$\Phi_O$ & CP-odd physical states for any (value of) $b_3$, $\Phi_O = G,~A$ or 
$\tilde\nu_-$ \\
\end{tabular}
\caption{\emph{Notation used in this paper for the various physical 
(mass-eigenstates) scalar states  and SU(2) (weak) scalar states.}}
\label{tab1}
\end{table}  

\section{Analytical derivation of cross-sections}

Let us denote the total cross-sections for $e^+e^- \to VV$ and
$e^+e^- \to t \bar t$ by $\sigma_V$ and $\sigma_t$
respectively. In the presence of the RPV SUSY interactions in
(\ref{lp}) and (\ref{bterm}), new $s$-channel scalar exchanges
have to be added at tree-level to the usual SM diagrams for these
processes. The interferences between the SM diagrams and our
$s$-channel scalar exchange diagrams (see Fig.~\ref{fig0}(e)) 
are $\propto m_e$ and are
therefore negligible. Thus, the total cross-sections above are
given by the simple sums $\sigma_V = \sigma_V^{SM} + \sigma_V^s$
and $\sigma_t = \sigma_t^{SM} + \sigma_t^s$, where $\sigma_V^s$
and $\sigma_t^s$ are those parts of the cross-sections
corresponding to the RPV scalar exchanges only:\footnote{Note that
in \cite{ourplb} we used the superscript $0$ to denote the scalar
exchange cross-sections while here the superscript $s$ is used
instead.}

\begin{eqnarray}
\sigma_V^s &\equiv& \sigma(e^+e^- \to \Phi_E \to VV)~,~ V=W~{\rm
or}~Z
\label{sigv0}~,\\
\sigma_t^s &\equiv& \sigma(e^+e^- \to \Phi_E,\Phi_O \to t \bar t)
\label{sigt0}~.
\end{eqnarray}

\n Recall that by $\Phi_E$ and $\Phi_O$ we mean  $\Phi_E = H,h$
and $\tilde\nu_+$ and $\Phi_O = A$ and $\tilde\nu_-$ (the
Goldstone boson G is not included) all of which has to be summed
in the corresponding amplitudes. Note again
that the CP-odd states $\Phi_O$ are absent in $\sigma_V^s$ since
they do not couple to $VV$ at tree-level.

In order to calculate $\sigma_V^s$ and $\sigma_t^s$ we need the
new Feynman rules for the vertices induced by the presence
of the RPV SUSY terms. The interactions  of the physical states
$\Phi_E$ and $\Phi_O$ are obtained by rotating the Feynman rules
of the RPC SUSY Lagrangian (see e.g., \cite{rosiek}) with the
matrices $S_E$ and $S_O$, respectively. Thus, if
$\Lambda_{\Phi^0_{E,\ell}}$ is an interaction vertex involving the
weak state $\Phi_{E,\ell}^0$ ($\ell=H^0,h^0$ or $\tilde\nu_+^0$),
then $\Lambda_{\Phi_{E,k}}$ - the vertex involving the physical
state $\Phi_{E,k}$ ($k=H,h$ or $\tilde\nu_+$) - is given by
$\Lambda_{\Phi_{E,k}} = S_E^{\ell k} \Lambda_{\Phi^0_{E,\ell}}$.
Similarly, in the CP-odd sector $\Lambda_{\Phi_{O,k}} = S_O^{\ell
k} \Lambda_{\Phi^0_{O,\ell}}$, where now $\ell=A^0$ or
$\tilde\nu_-^0$ and $k=A$ or $\tilde\nu_-$.

For the $\Phi_{E,k} VV$ coupling we then get:

\begin{equation}
\Lambda_{\Phi_{E,k} V_\mu V_\nu} = i (e/s_W) C_V m_V \left(c_\beta S^{1k}_E
+s_\beta S^{2k}_E \right) g_{\mu \nu}~,
\end{equation}

\n where $C_V=1(1/c_W)$ for $V=W(Z)$,
$s_W(c_W) \equiv \sin\theta_W(\cos\theta_W$)
and $c_\beta(s_\beta) \equiv \cos\beta(\sin\beta$).

The $\Phi_{E,k} t \bar t$ and $\Phi_{O,k} t \bar t$ couplings are:

\begin{eqnarray}
\Lambda_{\Phi_{E,k} t \bar t} &=& -\frac{i}{2} (e/s_W) \frac{m_t}{M_W s_\beta}
S_E^{2k}~,\\
\Lambda_{\Phi_{O,k} t \bar t} &=& -\frac{1}{2} (e/s_W)
\frac{m_t}{M_W s_\beta} S_O^{2k} \gamma_5~.
\end{eqnarray}

\n The $\Phi_{E,k} e^+e^-$ and $\Phi_{O,k} e^+e^-$ couplings are
obtained from the RPVTT term in (\ref{lp}). In particular,
neglecting the Higgs couplings to electrons, only the couplings
$\Lambda_{\tilde\nu_+^0 e^+ e^-} = i \lambda_{131}/\sqrt 2 $ and
$\Lambda_{\tilde\nu_-^0 e^+ e^-} = - \lambda_{131} \gamma_5/\sqrt
2 $ need to be rotated (recall that $\tilde\nu_\pm^0$
refers here only to the tau-sneutrino). We then get:

\begin{eqnarray}
\Lambda_{\Phi_{E,k} e^+ e^-} &=& \frac{i}{\sqrt 2} \lambda_{131} S_E^{3k}~,\\
\Lambda_{\Phi_{O,k} e^+ e^-} &=& -\frac{1}{\sqrt 2} \lambda_{131} S_O^{3k}
\gamma_5~.
\end{eqnarray}

\n The cross-sections in (\ref{sigv0}) and (\ref{sigt0}) are then
readily calculated in the new RPV interaction basis and are given
by:

\begin{eqnarray}
\sigma^s_V=\delta_V C_V^2 \frac{\alpha}{128 s_W^2} \frac{\beta_V
(3-2 \beta_V^2 +3 \beta_V^4)}{s (1-\beta_V^2)} \lambda_{131}^2
\times \mid \sum_{k=1}^3 S_E^{3k} A^k_V \hat\Pi^k_E \mid^2
\label{sigmav}~,
\end{eqnarray}

\n and

\begin{eqnarray}
\sigma^s_t = \frac{3 \alpha}{32 s_W^2} \left(\frac{m_t}{M_W
s_\beta}\right)^2 \frac{\beta_t}{s} \lambda_{131}^2 \left\{ \mid
\sum_{k=1}^3 S_O^{3k} S_O^{2k} \hat{\Pi}^k_O \mid^2 +
\beta_t^2 \mid \sum_{k=1}^3 S_E^{3k} S_E^{2k}
\hat{\Pi}^k_E \mid^2 \right\} \label{sigmat}~,
\end{eqnarray}

\n where $\delta_V=2(1)$ for $V=W(Z)$, $\beta_i =
\sqrt{1-4m_i^2/s}$ and $s$ is the square of the c.m. energy. We
have further defined the "reduced" $\Phi_{E}^k VV$ coupling:

\begin{equation}
A^k_V \equiv c_\beta S^{1k}_E +s_\beta S_E^{2k}  \label{av} ~,
\end{equation}

\n and the dimensionless propagator factors for the CP-even and
CP-odd scalars:

\begin{eqnarray}
\hat\Pi^k_{E,O} \equiv \left(1-(x^k_{E,O})^2+i x^k_{E,O} y^k_{E,O}
\right)^{-1}~,
\end{eqnarray}

\n where

\begin{eqnarray}
x^k_{E}; x^k_{O} \equiv \frac{m_{\Phi_{E,k}};m_{\Phi_{O,k}}}{\sqrt s}
~~,~~ y^k_{E};y^k_O
\equiv \frac{\Gamma_{\Phi_{E,k}};\Gamma_{\Phi_{O,k}}}{\sqrt s}
\label{gamsnu}~,
\end{eqnarray}

\n and $\Gamma_{\Phi_{E,k}}$($\Gamma_{\Phi_{O,k}}$) is the
$\Phi_{E,k}$($\Phi_{O,k}$) width.

As mentioned before, in our numerical results we will evaluate
$\sigma^s_V$ and $\sigma^s_t$ under the conditions of small RPV
($\varepsilon \ll 1$), a heavy Higgs spectrum ($(m_A^0)^2
 \gg M_Z^2$), $t_\beta^2 \gg 1$
and that no accidental mass degeneracy between scalars of the
same CP property occur. Under these assumptions it is possible to
give a qualitative description of the behavior of $\sigma^s_V$
and $\sigma^s_t$.

Let us consider first the cross-section $\sigma^s_V$. As was shown
in the previous section, when $\varepsilon \ll 1$ and $(m_A^0)^2
\gg M_Z^2$ we have $S_E^{11} \to s_\beta$ and $S_E^{21} \to -
c_\beta$ (see (\ref{seeps}) and (\ref{sedec})), which therefore
leads to $A^1_V \to 0$, where $A^1_V$ is the reduced $HVV$
coupling defined through (\ref{av}). Moreover, for $\varepsilon
\ll 1$, the element connecting $H$ to $\tilde\nu_+^0$ diminishes,
i.e., $S_E^{31} \ll 1$. Thus, the $H$ exchange contribution
to $\sigma_V^s$ ($\propto \Lambda_{He^+e^-} \times \Lambda_{HVV}
\sim S_E^{31} \times A^1_V$) is doubly suppressed.

As for the sneutrino-like state, $\tilde\nu_+$, due to
$\mid (M_E^2)_{13}/(M_E^2)_{23} \mid = t_\beta$ (see (\ref{me2})),
$\tilde\nu_+^0$ acquires a larger $\xi_d^0$ mixing (as compared
with $\xi_u^0$) which in turn implies a larger $H$ mixing,
since under the above conditions the $H$ mass-eigenstate is mostly
the $\xi_d$ weak-state (see (\ref{rel1}) and (\ref{rel2})).
Therefore, the $\tilde\nu_+$ couples to the gauge-bosons pair
mostly through its $H$ component and 
$\Lambda_{\tilde\nu_+ V V}$ and $\Lambda_{H V V}$ get comparable
(or equivalently $A^3_V \sim A^1_V$). On the other hand, for
$\varepsilon \ll 1$, $\tilde\nu_+$ has a much stronger (than $H$)
coupling to the incoming electron since it couples to $e^+e^-$
through its dominant $\tilde\nu_+^0$ component. In particular,
$S_E^{33} \to 1$ as $\varepsilon \to 0$. Thus, even though
$\Lambda_{\tilde\nu_+ V V} \sim \Lambda_{H V V}$, the
$\tilde\nu_+$ exchange contribution to $\sigma_V^s$, being
$\propto S_E^{33} \times A^3_V$, will be much more pronounced than
the $H$ one, due to $S^{33}_E \gg S^{31}_E$.

Therefore, the more favorite scenario for observing such a
sneutrino-Higgs mixing resonance in $VV$ pair production is when
the $\tilde\nu_+$ resonates. Note that a light Higgs ($h$)
resonance in on-shell $VV$ pair production is theoretically
excluded, since the c.m. energy required to produce an on-shell
$VV$ pair is at least $\sim 25$ GeV above the highest possible
$m_h$ (the theoretical upper limit on $m_h$ is $\sim 135$ GeV).
Since we are only interested in the case of a
sneutrino--Higgs resonance enhancement in $\sigma_V^s$, the $h$
contribution which is always ``far'' from resonance is negligible
- in particular near the $\tilde\nu_+$ resonance.

The case of $s$-channel scalar exchanges in $e^+e^- \to t \bar t$
is a little more complicated due to the extra CP-odd scalar
exchanges in the $s$-channel. In the limit $(m_A^0)^2 \gg M_Z^2$
and $\varepsilon \ll 1$, we have $m_A \sim m_H$ and also $m_{s
\nu}^+ \sim m_{s \nu}^-$ (see previous section). Therefore, if the
$\tilde\nu_+$ resonates then necessarily also the $\tilde\nu_-$
will be close to resonance and is expected to yield a
comparable enhancement in the vicinity of a $\tilde\nu_+$
resonance. Similarly, a $H$ near-resonance enhancement will
necessarily follow from an $A$ resonance. 

As it turns out, the situation here is similar 
to that in $\sigma_V^s$
since here also the sneutrino-like states, $\tilde\nu_+$ and $\tilde\nu_-$, 
will potentially yield a stronger resonance
enhancement than the Higgs-like states, $H$ and $A$. This can
be understood as follows. From (\ref{sigmat}) we see that, apart
from the common factors that enter $\sigma_t^s$ for each of the scalar states
exchanges, the relative strength between the $A$ and
$\tilde\nu_-$ contributions as well as between the $H$ and the 
$\tilde\nu_+$ ones 
are determined by the quantities
$S_O^{3k} \times S_O^{2k}$ and $S_E^{3k} \times S_E^{2k}$, respectively. 
Considering the cases $k=1$ (the $A$ and $H$ 
exchanges) and $k=3$ (the $\tilde\nu_-$ and $\tilde\nu_+$ exchanges), 
we have (see (\ref{seeps}), (\ref{soeps}) and (\ref{sedec})):

\begin{eqnarray}
&& S_O^{31} ~\stackrel{\varepsilon \to 0} {\longrightarrow}~ 0 ~~,~~ 
S_O^{33} ~\stackrel{\varepsilon \to 0} {\longrightarrow}~ 1 \\
&& S_E^{31} ~\stackrel{\varepsilon \to 0} {\longrightarrow}~ 0 ~~,~~ 
S_E^{33} ~\stackrel{\varepsilon \to 0} {\longrightarrow}~ 1~,
\end{eqnarray}

\n and

\begin{eqnarray}
&& S_O^{21} ~\stackrel{\varepsilon \to 0} {\longrightarrow}~
c_\beta ~\stackrel{t_\beta^2 \gg 1} {\longrightarrow}~ 0 ~~,~~
S_O^{23} ~\stackrel{\varepsilon \to 0} {\longrightarrow}~ 0 \\
&& S_E^{21} ~\stackrel{\varepsilon \to 0} 
{\longrightarrow}~ s_\alpha 
~\stackrel{m_A^2 \gg M_Z^2} {\longrightarrow}~
- c_\beta ~\stackrel{t_\beta^2 \gg 1} {\longrightarrow}~ 0 ~~,~~
S_E^{23} ~\stackrel{\varepsilon \to 0} {\longrightarrow}~ 0~.
\end{eqnarray}

\n Therefore, we see that also in the case of $\sigma_t^s$ the
sneutrino-like exchanges may potentially yield a
stronger resonance when $\varepsilon \ll 1$, $m_A^2 \gg M_Z^2$ 
and $t_\beta^2 \gg 1$
since in this limit $S_O^{33} \times S_O^{23} \gg S_O^{31} \times
S_O^{21}$ and $S_E^{33} \times S_E^{23} \gg S_E^{31} \times
S_E^{21}$, mainly due to $S_O^{33} \gg S_O^{31}$ and 
$S_E^{33} \gg S_E^{31}$, respectively. Hence, since we
are interested in the largest possible resonance enhancement in both
$e^+ e^- \to VV$ and $e^+e^- \to t \bar t$, in our numerical
analysis we will investigate only the cases of 
resonances emanating from exchanges of the sneutrino-like
admixtures while setting the masses of the Higgs-like states 
to be sufficiently away from the c.m. energy of the collider.  
In particular, in the $VV$
production we will consider a resonance caused by the CP-even
sneutrino-like state, $\tilde\nu_+$, and in $t\bar t$ production
we will investigate the ``combined'' resonance effect 
that may emerge from the
 CP-odd and CP-even sneutrino states, $\tilde\nu_-$ and $\tilde\nu_+$. 
It should be
clear, however, that if $m_A \sim m_{s \nu}^\pm$ (that will be the
case for $\varepsilon \ll 1$ and if $m_A^0 \sim m_{s \nu}^0$),
then both $\sigma_V^s$ and $\sigma_t^s$ may be further enhanced
since there will be several scalar states whose masses are nearly
degenerate and happen to lie close to the c.m. energy. 
In particular, under the conditions $(m_A^0)^2 \gg
M_Z^2$ and $\varepsilon \ll 1$, choosing $m_A^0 \sim m_{s \nu}^0$
implies $m_{s \nu}^+ \sim m_{s \nu}^- \sim m_H \sim m_A$ in which
case $\sigma_V^s$ may exhibit a "double" resonance enhancement and
$\sigma_t^s$ may have a "four-fold" resonant structure. Here,
we will not consider such a possibility of an accidental
mass degeneracy among the scalar states involved which may give
rise to these multi-resonant structures in $\sigma_V^s$ and in
$\sigma_t^s$.

Finally, the $\tilde\nu_\pm$ widths ($\Gamma_{\tilde\nu_\pm}$ in
(\ref{gamsnu})) need to be included, since they crucially control the
behavior of $\sigma_V^s$ and of $\sigma_t^s$ in the vicinity of
our $\tilde\nu_+$ and $\tilde\nu_-$ resonances.
Assuming that the lightest neutralino ($\tilde\chi_1^0$) is the
Lightest SUSY Particle (LSP) and also that $m_{s \nu}^\pm >
m_{\tilde\chi_1^+}$, where $\tilde\chi_1^+$ is the lighter
chargino, then the RPC two-body decays $\tilde\nu_\pm \to
\tilde\chi_1^0 \nu_\tau,~ \tilde\chi_1^+ \tau$ are open and
dominate. Indeed, for $m_A^2 \gg M_Z^2$ and following the
traditional assumption of an underlying grand unification with a
common gaugino mass parameter $m_{1/2} < m_{s \nu}^0$, the mass
hierarchy $m_{\tilde\chi_1^0} < m_{\tilde\chi_2^0} \sim
m_{\tilde\chi_1^+} < m_{s \nu}^0$ and $m_{\tilde\chi_{3,4}^0} \sim
m_{\tilde\chi_2^+} > m_{s \nu}^0$ is possible, e.g., when $m_{s
\nu}^\pm < m_A$ \cite{djuadi} (recall that in the case of
interest, $\varepsilon \ll 1$, we have $m_{s \nu}^0 \sim m_{s
\nu}^+ \sim m_{s \nu}^-$, see previous section). Thus, upon
ignoring phase space factors, a viable conservative estimate is
(see e.g., Barger {\it et al.} in \cite{snures} and
\cite{ourcpsnu}): $\Gamma_{\tilde\nu_\pm} \sim \Gamma
(\tilde\nu_\pm \to \tilde\chi_{1,2}^0 \nu_\tau) + \Gamma
(\tilde\nu_\pm \to \tilde\chi_1^+ \tau) \sim 10^{-2} m_{s
\nu}^\pm$ which we use below (for the ranges of
$\varepsilon,~m_A^0$ and $m_{s \nu}^0$ considered the possible RPV
decays are sufficiently smaller and $\Gamma_{\tilde\nu_\pm} \sim
\Gamma_{\tilde\nu_\pm^0}$ since $S_E^{33},~S_O^{33} \to 1$). Also,
for reasons explained above, $\Gamma_H$, $\Gamma_A$ and $\Gamma_h$
have a negligible effect on our $\tilde\nu_\pm$ resonances and are
therefore neglected.

\section{Sneutrino-like resonance in 
$\lowercase{e}^+ \lowercase{e}^- \to VV$. Numerical results}

Before presenting our numerical results for $\sigma_V^s$ we wish
to note the following:

\begin{itemize}

\item Sufficiently away from threshold (which occurs at $\beta_V \to 1$),
$\sigma_W^s/\sigma_Z^s \sim (\delta_W c_W^2/\delta_Z) \times
(M_Z/M_W)^2 \sim 2$ and, since typically
$\sigma_W^{SM}/\sigma_Z^{SM} > 10$, the relative effect of the
scalar exchange cross-section is more pronounced in the $ZZ$
channel. Therefore, below we will present results mainly for the
$ZZ$ production case (for $\sigma_Z^s$). It should be kept in mind, however,
that $\sigma_W^s$ is roughly a factor of 2 larger than
$\sigma_Z^s$ and exhibits the same behavior as a function of the
relevant RPV parameter space.

\item As mentioned in the previous section, for $\varepsilon \ll 1$ and
when $(m_A^0)^2 \gg m_Z^2$, we have $\Lambda_{hVV} \to 1$ and
$\Lambda_{HVV} \to 0$. If in addition $t_\beta^2 \gg 1$, one has
$\xi_d^0 \to H$ and $\xi_u^0 \to h$ so that in conjunction with
$\left(\Lambda_{h VV}/\Lambda_{H VV} \right) \gg 1$ also gives
$\left(\Lambda_{\xi_u^0 VV}/\Lambda_{\xi_d^0 VV} \right) \gg 1$.
Therefore, since the $\tilde\nu_+^0 - \xi_u^0$ mixing decreases
with $\tan\beta$ [i.e., $\mid (M_E^2)_{23} \mid = b_3 = \varepsilon
(m_A^0)^2/t_\beta$, see (\ref{me2})], as $t_\beta$ increases the
sneutrino ``prefers'' to mix more with $\xi_d^0$ which has a
suppressed coupling to $VV$ in this limit. As a consequence, the
$\tilde\nu_+$ resonance effect in $\sigma_V^s$ drops with
$\tan\beta$ in the limit of small RPV and when $m_A^2 \gg M_Z^2$.
In what follows we will choose the two values $t_\beta
=3$ and $t_\beta=50$ representing low and high $\tan\beta$
scenarios, respectively (note that already with our low
$\tan\beta$ value, $t_\beta=3$, $t_\beta^2$ is about an order of
magnitude larger than unity). Following the above reasoning, it
should be clear from the outset that $\sigma_V^s(t_\beta=3) \gg
\sigma_V^s(t_\beta=50)$.

\end{itemize}

\begin{figure}[htb]
\psfull
 \begin{center}
  \leavevmode
  \epsfig{file=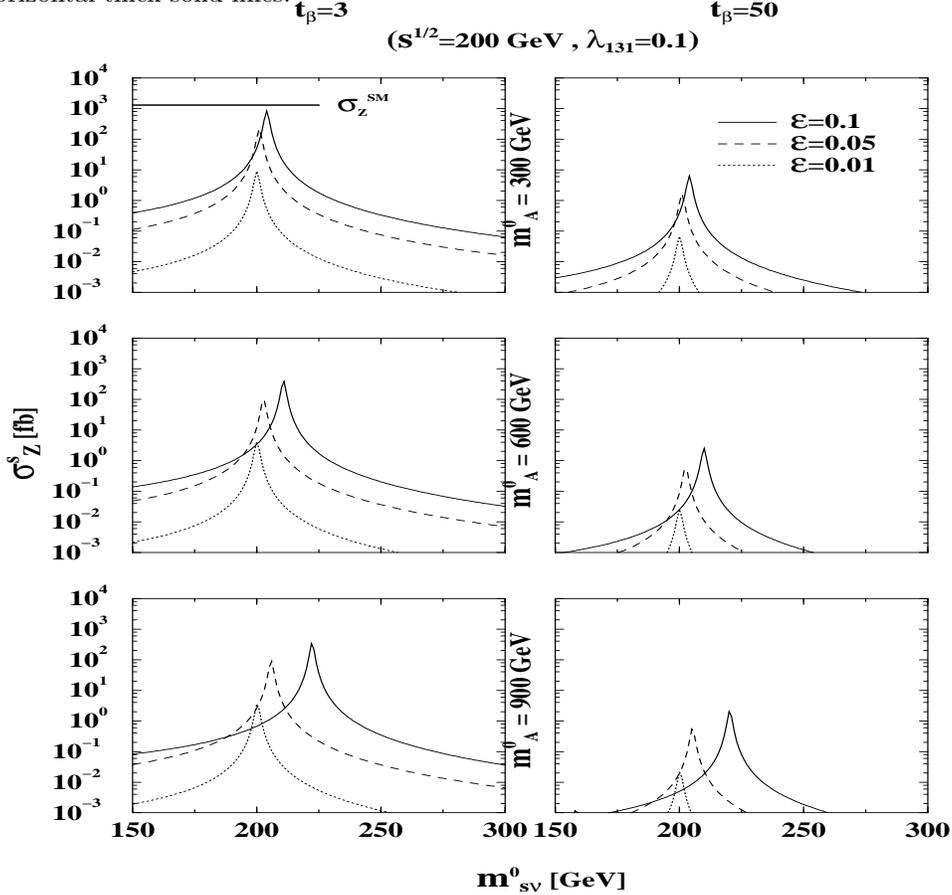,height=11cm,width=13cm,bbllx=0cm,bblly=1cm,bburx=20cm,bbury=25cm,angle=0}
 \end{center}
\caption{\emph{$\sigma_Z^s$ as a function of $m_{s \nu}^0$, for
$m_A^0=300,~600$ and $900$ GeV, for $t_\beta=3$ (left figures) and
$t_\beta=50$ (right figures). For all combinations of $m_A^0$ and
$t_\beta$ values, $\sigma_Z^s$ is shown for a c.m. energy of $\sqrt
s=200$ GeV with $\varepsilon=0.1,~0.05$ and $0.01$. Also, 
$\lambda_{131}=0.1$ is used (recall that $\sigma_Z^s$ scales as
$\lambda_{131}^2$). The SM $ZZ$ cross-sections for $\sqrt s=200$
is also shown by the horizontal thick solid line.}} \label{fig3}
\end{figure}

\begin{figure}[htb]
\psfull
 \begin{center}
  \leavevmode
  \epsfig{file=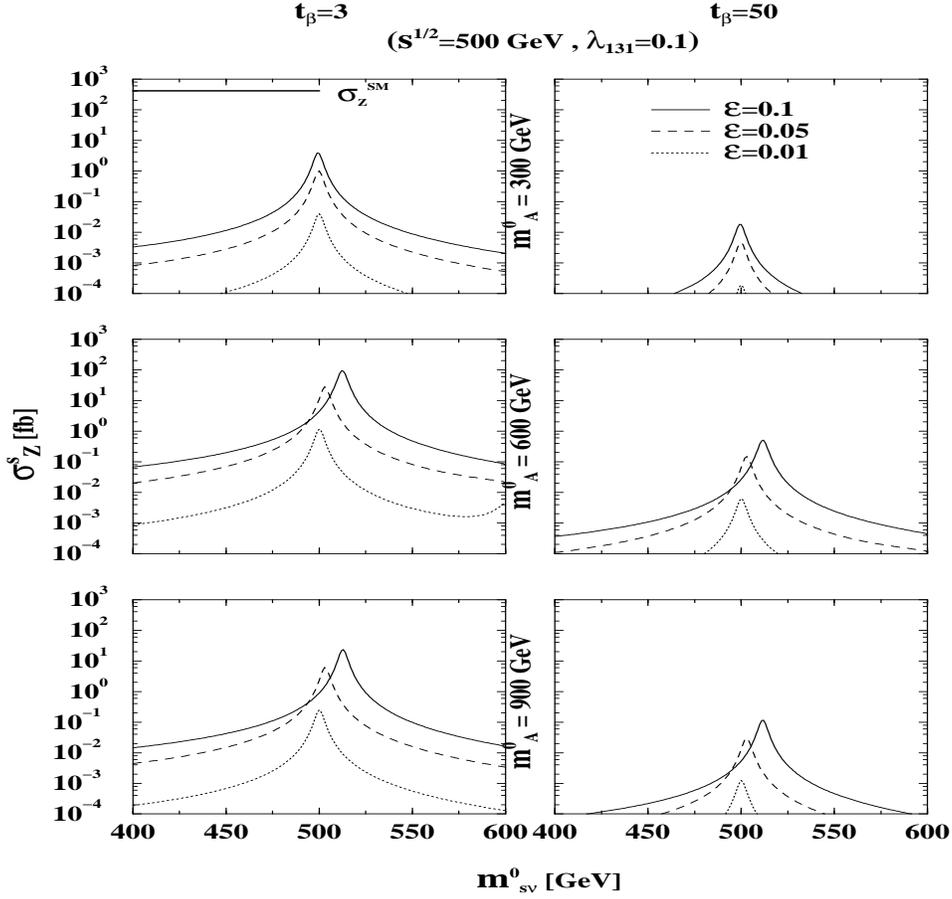,height=11cm,width=13cm,bbllx=0cm,bblly=1cm,bburx=20cm,bbury=25cm,angle=0}
 \end{center}
\caption{\emph{Same as Fig.~\ref{fig3} but for a c.m. energy of
500 GeV.}}
\label{fig4}
\end{figure}

In Figs.~\ref{fig3} and \ref{fig4} we show $\sigma_Z^s$ as a
function of $m_{s \nu}^0$ for c.m. energies of $\sqrt s=200$ and
500 GeV, respectively. This is shown for $t_\beta=3,~50$ and for
$m_A^0=300,~600,~900$ GeV. For definiteness we choose
$\varepsilon=0.01,~0.05$ or $0.1$ and
$\lambda_{131}=0.1$.\footnote{$\sigma_V^s$ is insensitive to the
signs of $\varepsilon$ and $\lambda_{131}$.} The SM cross-sections
$\sigma_Z^{SM}(\sqrt s=200~{\rm GeV}) \sim 1.29$ [pb] and
$\sigma_Z^{SM}(\sqrt s=500~{\rm GeV}) \sim 0.41$ [pb] are also
shown by the horizontal thick solid lines.


As expected, $\sigma_Z^s$ is larger for a smaller $|m_A^0 - m_{s
\nu}^0|$ mass splitting since the sneutrino--Higgs mixing
phenomena is proportional to $[(m_A^0)^2 - (m_{s
\nu}^0)^2]^{-1}$ (see (\ref{shift1})).
Clearly, the scalar exchange cross-section can be noticeable and
statistically significant within some interval of $m_{s \nu}^+$
around the c.m. energy. As we shall see below, the interval $|m_{s
\nu}^+ - \sqrt s|$ for which the RPV signal is statistically
significant may range from a few GeV to a few tens of GeV
depending on $\varepsilon$ and the rest of the SUSY parameter
space involved.

\begin{figure}[htb]
\psfull
 \begin{center}
  \leavevmode
  \epsfig{file=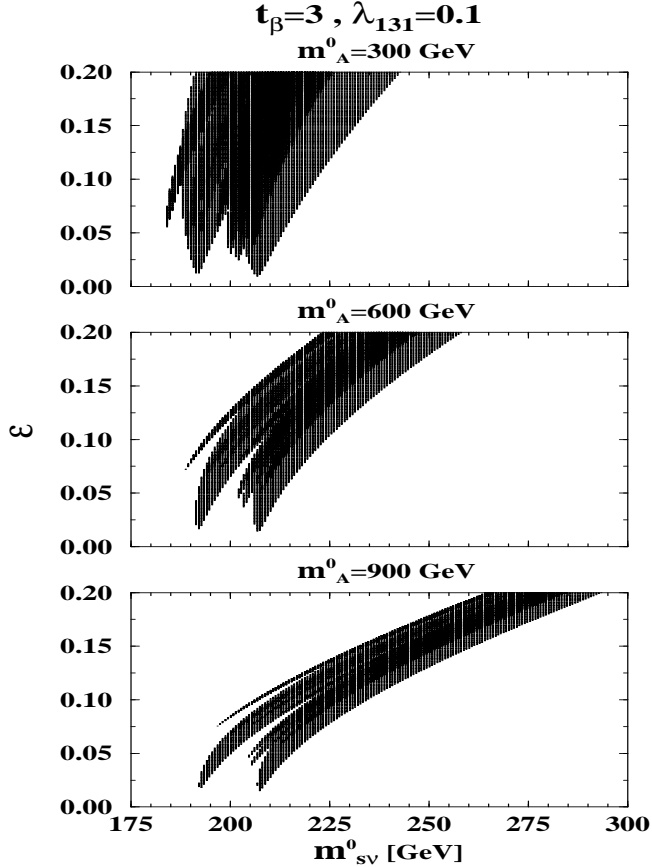,height=11cm,width=13cm,bbllx=0cm,bblly=1cm,bburx=20cm,bbury=25cm,angle=0}
 \end{center}
\caption{\emph{$1 \sigma$ excluded regions in the $\varepsilon -
m_{s \nu}^0$ plane from the LEP2 measurements of the $WW$ and $ZZ$
cross-sections (see text), for $t_\beta=3$ with $m_A^0=300,~600$
or $900$ GeV and for $\lambda_{131}=0.1$. Note that the fingers like
shape of the shaded area is an artifact of the discrete set of
c.m. energies used to derive these limits in accordance with the
LEP2 runs.}} \label{fig5}
\end{figure}

Let us first consider the case of $\sqrt s$ around 200 GeV - that of
LEP2 energies. We can use the measured values of the $WW$ and $ZZ$
cross-sections at LEP2 to place further bounds on the $\varepsilon
- m_{s \nu}^0$ plane for a given $m_A^0$ and $t_\beta$. This is
shown in Fig.~\ref{fig5} where we have taken the measured cross-sections
$\sigma_Z^{exp}$ and $\sigma_W^{exp}$ (combined from the 4 LEP experiments)
from the 183, 189, 192, 196, 200, 202, 205 and 207 GeV LEP2 runs
as given in \cite{lep2sigs}. In particular, for each run we take
the experimentally measured ($\sigma_V^{exp} \pm \Delta \sigma_V^{exp}$)
and the SM theoretical ($\sigma_V^{SM} \pm \Delta \sigma_V^{SM}$)
cross-sections (also given in
\cite{lep2sigs}),\footnote{For the $ZZ$ and $WW$ SM cross-sections
we use the results of the ZZTO and YFSWW3 Monte-Carlos,
respectively, where we take a $2\%$ theoretical error for the ZZTO
prediction and no error for the YFSWW3 one, see \cite{lep2sigs}.}
and require
that:\footnote{Since $\sigma_V^s$ always increases the SM 
cross-section, we do not include the cases in which
$(\sigma_V^{exp} - \sigma_V^{SM}) + \sqrt{(\Delta
\sigma_V^{exp})^2+(\Delta \sigma_V^{SM})^2} <0$.}

\begin{eqnarray}
\sigma_V^s < (\sigma_V^{exp} - \sigma_V^{SM}) + \sqrt{(\Delta
\sigma_V^{exp})^2+(\Delta \sigma_V^{SM})^2} \label{siglimit} ~.
\end{eqnarray}

\n The $1\sigma$ excluded regions in Fig.~\ref{fig5} are derived
through (\ref{siglimit}) for $\lambda_{131}=0.1$ and for
$t_\beta=3$ with $m_A^0=300,~600$ or 900 GeV. For $t_\beta=50$ and
$m_A^0 \leq 300$ GeV we find that no such limits can be imposed
since the RPV cross-sections $\sigma_V^s$ are too small for such a
large $t_\beta$ value (we find a ``tiny'' excluded area in the
$\varepsilon - m_{s \nu}^0$ plane for $t_\beta=50$ and $m_A^0 =
300$ GeV and when $\varepsilon \gsim 0.15$).

Evidently, the limits coming from the $ZZ$ and $WW$ cross-sections
measurements give further restrictions for $t_\beta=3$
at low $\varepsilon$ values
(below $\sim 0.2$), in a sneutrino mass range of several tens of
GeV\footnote{Note that, since $b_3 = \varepsilon
(m_A^0)^2/t_\beta$, these $1 \sigma$ limits can be directly
translated into limits on the $b_3 - m_{s \nu}^0$ plane.}, for
which there are no bounds coming from the LEP2 limits on $m_h$
(see Fig.~\ref{fig2}). Note that the fingers like shape of the
shaded area in Fig.~\ref{fig5} is an artifact coming from the 
discrete set of c.m. energies used in accordance with the
LEP2 runs.

Let us now examine the sensitivity of 
a future 500 GeV $e^+e^-$ collider to the RPV sneutrino-like
resonance effect in $e^+e^- \to VV$. We will require that our new
RPV cross-section signal be smaller than the experimental error as
in (\ref{siglimit}), where now all cross-sections are for a c.m.
energy of 500 GeV. We will
assume that the central value of the future measured cross-section
for $VV$ production at a c.m. energy of 500 GeV
($\sigma_V^{exp}$) coincides with the corresponding SM theoretical
value, i.e., $\sigma_V^{exp} = \sigma_V^{SM}$. Also, we combine
the theoretical ($\Delta \sigma_t^{SM}$) and experimental ($\Delta
\sigma_V^{exp}$) errors and
scale it with the measured cross-section as follows:

\begin{equation}
\sqrt{(\Delta \sigma_V^{exp})^2+(\Delta \sigma_V^{SM})^2} \equiv
\sigma_V^{exp} \delta_V^\sigma = \sigma_V^{SM} \delta_V^\sigma 
\label{deltav}~,
\end{equation}

\n such that $\delta_V^\sigma$ now represents the overall
(statistical + systematic + theoretical) error per event [e.g., if
$\Delta\sigma_V^{SM} \ll \Delta\sigma_V^{exp}$, then
$\delta_V^\sigma=0.1$ corresponds to a $10\%$ error in the actual
measurement of $\sigma(e^+e^- \to VV)$]. Thus, the condition
for the observability of $\sigma_V^s$ reduces to (at the $1\sigma$ level):

\begin{eqnarray}
\frac{\sigma_V^s}{\sigma_V^{SM}} > \delta_V^\sigma
\label{siglimitv}~.
\end{eqnarray}

\noindent Using (\ref{siglimitv}), we can calculate
the sneutrino-like mass range
for which its contribution to the $WW$ and $ZZ$ cross-sections
becomes observable (at $1\sigma$). For example, if $\sigma(e^+e^- \to ZZ)$ 
is measured at a 500 GeV collider with an overall 
$5\%$ or $10\%$ error (i.e., $\delta_Z^\sigma=0.05$ or $0.1$), then,
for $m_A^0=600$, $t_\beta=3$ and $\varepsilon=\lambda_{131}=0.1$, 
the scalar resonance
signal in $ZZ$ production will be observable within the sneutrino-like
mass intervals $495~{\rm GeV} \lsim m_{s \nu}^+ \lsim 504~{\rm GeV}$ or
$497~{\rm GeV} \lsim m_{s \nu}^+ \lsim 502~{\rm GeV}$, respectively.    
These mass ranges are further enlarged if an angular
cut on the c.m. scattering angle, $\theta$, is imposed, due to 
the different angular dependence of the signal and SM background. For
example, with $-0.5 \lsim \cos\theta \lsim 0.5$, we find that 
the corresponding mass intervals are 
$490~{\rm GeV} \lsim m_{s \nu}^+ \lsim 509~{\rm GeV}$ for 
$\delta_Z^\sigma=0.05$ and 
$494~{\rm GeV} \lsim m_{s \nu}^+ \lsim 507~{\rm GeV}$ for 
$\delta_Z^\sigma=0.1$.

For the $WW$ production case, with the angular cut 
$0 \lsim \cos\theta \lsim 1$ and 
for the same values of $t_\beta$, $m_A^0$, $\varepsilon$ and
$\lambda_{131}$ as above, 
the corresponding mass
intervals are
$494~{\rm GeV} \lsim m_{s \nu}^+ \lsim 506~{\rm GeV}$ for 
$\delta_W^\sigma=0.05$ and 
$496~{\rm GeV} \lsim m_{s \nu}^+ \lsim 504~{\rm GeV}$ for 
$\delta_W^\sigma=0.1$.

In the next section we will show that a much stronger RPV
scalar resonance enhancement is expected
in the reaction $e^+e^- \to t \bar t$.

\section{Sneutrino-like resonance in 
$\lowercase{e}^+ \lowercase{e}^- \to \lowercase{t} \bar{\lowercase{t}}$. Numerical results}

As explained in section 3, We will mainly
focus below on the case of a combined $\tilde\nu_- - \tilde\nu_+$ 
resonance in
$\sigma_t^s$ since those are expected to yield
the largest possible scalar-resonance
signal in $e^+e^- \to t \bar t$.
Recall that $\varepsilon \ll 1$ implies a small
$\tilde\nu_- - \tilde\nu_+$ mixing so that if one of 
the sneutrino-like states resonates so does the other which has an opposite 
CP property.  

The $\tilde\nu_-$ and $\tilde\nu_+$ states couple 
to the top-quark through their
$\phi_u^0$ and $\xi_u^0$ components (the CP-odd and CP-even 
$H_u^0$ states, respectively, see
(\ref{components})). Since the RPV $\tilde\nu_-^0 -
\phi_u^0$ and $\tilde\nu_+^0 -
\xi_u^0$ mixings are proportional to $ b_3 = \varepsilon (m_A^0)^2 /
t_\beta$ (see $(M_O^2)_{23}$ and $(M_E^2)_{23}$ in (\ref{mo2}) 
and (\ref{me2})) and since the
$\phi_u^0 t \bar t$ and $\xi_u^0 t \bar t$ couplings themselves also 
go like $1/t_\beta$,
the $\tilde\nu_- t \bar t$ and $\tilde\nu_+ t \bar t$ couplings drop 
with $t_\beta$ and we
expect $\sigma_t^s$ to significantly decrease as $t_\beta$ is
increased. Nonetheless, in most instances below we will present
our numerical results again for the two values $t_\beta=3$ and
$t_\beta=50$ in order to illustrate this behavior of $\sigma_t^s$
as a function of $t_\beta$.

\begin{figure}[htb]
\psfull
 \begin{center}
  \leavevmode
  \epsfig{file=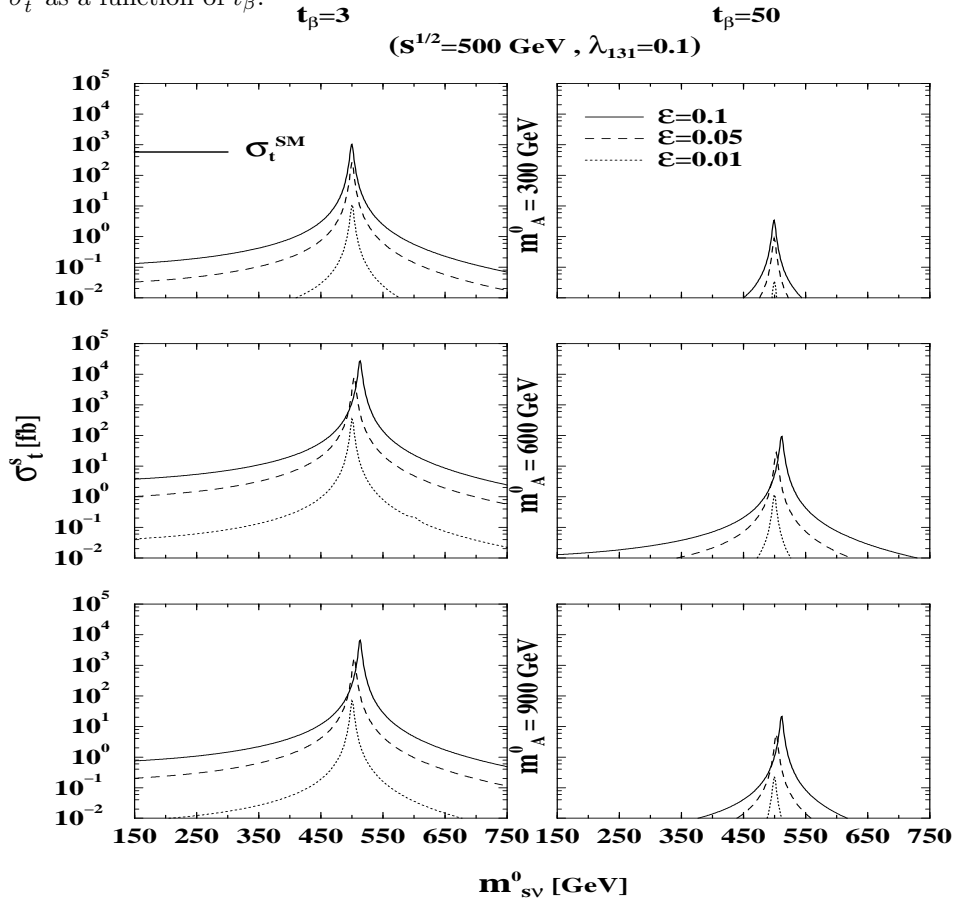,height=11cm,width=13cm,bbllx=0cm,bblly=1cm,bburx=20cm,bbury=25cm,angle=0}
 \end{center}
\caption{\emph{$\sigma_t^s$ as a function of $m_{s \nu}^0$, for
$m_A^0=300,~600$ and $900$ GeV, for $t_\beta=3$ (left figures) and
$t_\beta=50$ (right figures). For all combinations of $m_A^0$ and
$t_\beta$ values, $\sigma_t^s$ is shown for a c.m. energy of $\sqrt
s=500$ and with $\varepsilon=0.1,~0.05$ and $0.01$. Also,
$\lambda_{131}=0.1$ is used (recall that $\sigma_t^s$ scales as
$\lambda_{131}^2$). The SM $t \bar t$ cross-section for $\sqrt
s=500$ is also shown by the horizontal thick solid line.}}
\label{fig6}
\end{figure}
\begin{figure}[htb]
\psfull
 \begin{center}
  \leavevmode
  \epsfig{file=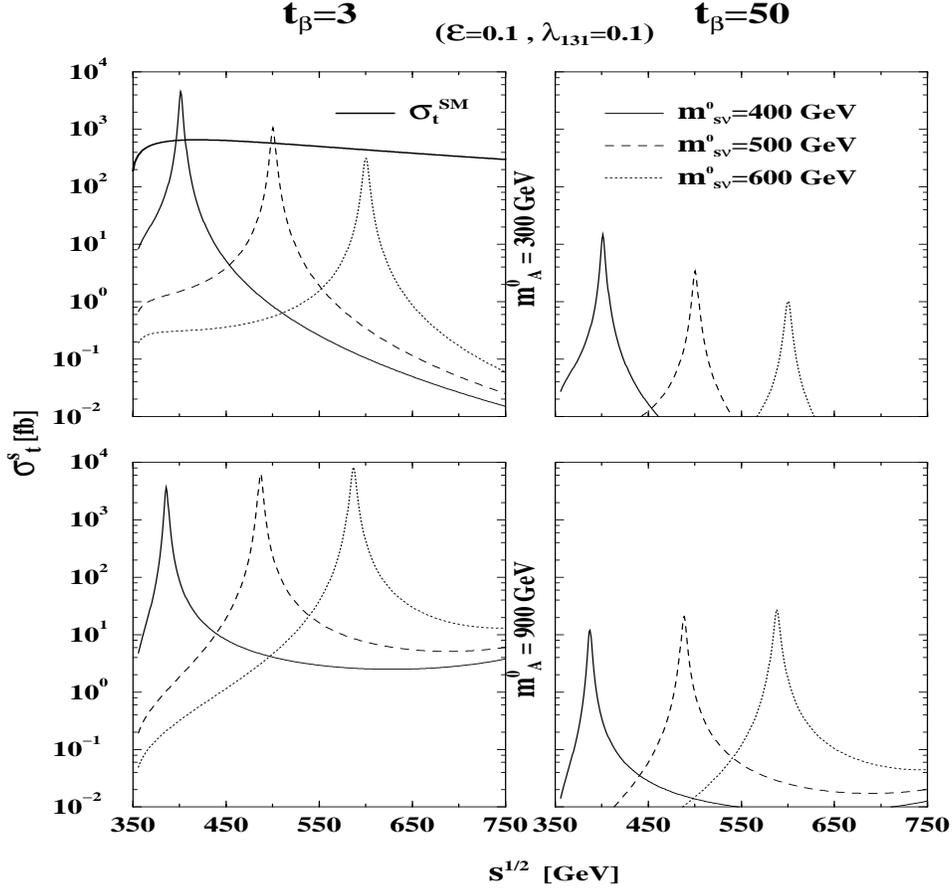,height=11cm,width=13cm,bbllx=0cm,bblly=1cm,bburx=20cm,bbury=25cm,angle=0}
 \end{center}
\caption{\emph{$\sigma_t^s$ as a function of the c.m. energy
$\sqrt s$, for $m_A^0=300$ and $900$ GeV and for $t_\beta=3$ (left
figures) or $t_\beta=50$ (right figures). For all combinations of
$m_A^0$ and $t_\beta$ values $\sigma_t^s$ is shown for
$\varepsilon=0.1=\lambda_{131}=0.1$ and for $m_{s \nu}^0=400,~500$
and $600$ GeV. The SM $t \bar t$ cross-section is also shown by the
thick solid line. See also caption to Fig.~\ref{fig6}.}}
\label{fig7}
\end{figure}

In Figs.~\ref{fig6} we plot $\sigma_t^s$ as a function of $m_{s
\nu}^0$ for an $e^+e^-$ collider with a c.m. energy of 500 GeV.
This is shown for $t_\beta=3,~50$ and for $m_A^0=300,~600$ or 900
GeV. The RPV couplings are set to $\varepsilon=0.01,~0.05$ or
$0.1$ and $\lambda_{131}=0.1$.\footnote{$\sigma_t^s$ is also
insensitive to the signs of $\varepsilon$ and $\lambda_{131}$.}
The SM cross-section $\sigma_t^{SM}(\sqrt s=500~{\rm GeV}) \sim
580$ [fb] is also shown by the horizontal thick solid line.

In Figs.~\ref{fig7} we show $\sigma_t^s$ as a function of the c.m.
energy, $\sqrt s$, ranging from the $t \bar t$ threshold to 750 GeV. 
Here we fixed the RPV couplings to be $\varepsilon=
\lambda_{131}=0.1$ and we depicted $\sigma_t^s$ for combinations
of $t_\beta=3,~50$ with $m_A^0=300$ or 900 GeV, where our input
sneutrino mass (i.e., in the RPC limit) was given three values:
$m_{s \nu}^0 = 400,~ 500$ and 600 GeV. The SM $t \bar t$
production cross-section, $\sigma_t^{SM}$ is again shown by a
thick solid line.

From Figs.~\ref{fig6} and \ref{fig7} it is evident that the scalar
exchange cross-section $\sigma_t^s$ decreases significantly as
$t_\beta$ is increased. Also here, as expected, $\sigma_t^s$
is larger for a smaller $|m_A^0 - m_{s \nu}^0|$ mass splitting,
due to factors of $[(m_A^0)^2 - (m_{s \nu}^0)^2]^{-1}$ which enter
the mixing matrices $S_O$ and $S_E$ in $\sigma_t^s$ (see section 3).

Evidently, the scalar exchange contribution in $e^+e^- \to t \bar t$
can be statistically significant within a rather large mass range
of $m_{s \nu}^-$ and $m_{s \nu}^+$ around resonance. As we shall see below, 
the
interval $|m_{s \nu}^- - \sqrt s|$, for which this RPV resonance
signal may be observable in $t \bar t$ production in a future
$e^+e^-$ high energy collider can range from a few tens of GeV
up to more than a hundred GeV, depending on theoretical parameters
such as $\varepsilon, \lambda_{131}$
and also
on the precision that will be achieved in measuring observable
quantities.

Let us investigate more quantitavely the limits that can be placed on
this RPV scalar mixing scenario in case that no such resonant
enhancement will be detected in $e^+e^- \to t \bar t$ at a 500 GeV
$e^+e^-$ machine. To estimate that, we require again that our new
RPV cross-section signal be smaller than the experimental error as
in (\ref{siglimit}). Here also, we
assume that the central value of the future measured cross-section
for $t \bar t$ production at a c.m. energy of 500 GeV
($\sigma_t^{exp}$) coincides with the corresponding SM theoretical
value, i.e., $\sigma_t^{exp} = \sigma_t^{SM}$, and we combine
the theoretical ($\Delta \sigma_t^{SM}$) and experimental ($\Delta
\sigma_t^{exp}$) errors to
scale with the measured cross-section as in (\ref{deltav}).
%
%\begin{equation}
%\sqrt{(\Delta \sigma_t^{exp})^2+(\Delta \sigma_t^{SM})^2} \equiv
%\sigma_t^{exp} \delta_t^\sigma = \sigma_t^{SM} \delta_t^\sigma ~,
%\end{equation}
%
Then, the condition
for observability of $\sigma_t^s$ (at the $1\sigma$ level) becomes 
(see also (\ref{siglimitv})):

\begin{eqnarray}
\frac{\sigma_t^s}{\sigma_t^{SM}} > \delta_t^\sigma
\label{siglimittop}~.
\end{eqnarray}

\n As an example, using (\ref{siglimittop}) we find that if the $t
\bar t$ production cross-section is measured with an overall
$10\%$ error ($\delta_t^\sigma=0.1$), then, for $t_\beta=3$ and
$\varepsilon=\lambda_{131}=0.1$, one can exclude the
sneutrino mass intervals $492 ~ {\rm GeV}~ \lsim m_{s \nu}^0
\lsim 507 ~{\rm GeV}$, $463 ~ {\rm GeV}~ \lsim m_{s \nu}^0 \lsim
560~{\rm GeV} $ and $492 ~ {\rm GeV}~ \lsim m_{s \nu}^0 \lsim 534
~{\rm GeV}$ for $m_A^0 =300,~600$ and 900 GeV, respectively. In
terms of the masses of the physical states $\tilde\nu_+$ and
$\tilde\nu_-$ (recall that $m_{s \nu}^- \sim m_{s \nu}^+$ for
$\varepsilon=0.1$, see Fig.~\ref{fig1}) the above excluded mass
ranges remain roughly the same, however, centered around $\sim
500$ GeV.

Asymmetries are often better probes of new physics
since they involve ratios of cross-sections. From the experimental point of
view, ratios of cross-sections can be determined with larger accuracy
since their systematic errors are usually in
better control compared to ``simple''
cross-sections measurements. We therefore expect asymmetries such as the
Forward-Backward (FB) asymmetry:

\begin{eqnarray}
A_{FB} \equiv \frac{\int_0^{\pi/2} \left\{ d\sigma(\theta) -
d\sigma(\pi -\theta) \right\}}{\sigma} \label{fbasym}~,
\end{eqnarray}

\n to be more sensitive to our new RPV scalar resonance effect in
$e^+e^- \to t \bar t$ due to the better accuracy with which it can be
measured.

In general, an $s$-channel scalar exchange does not have a FB
asymmetry since the corresponding cross-section (in our case
$\sigma_t^s$ or $\sigma_V^s$) does not depend on $\theta$. In any given
process for which there exists a non-vanishing FB asymmetry within
the SM, such a scalar exchange will reduce its SM (absolute) value
since it will make no contribution to the numerator of
(\ref{fbasym}) while increasing the total cross-section and
therefore the denominator in (\ref{fbasym}). 
In top pair production various types of asymmetries can be
constructed, e.g., polarization asymmetries, FB asymmetries and also
combinations of these two. Here we consider only the simplest FB
asymmetry for unpolarized top quarks and unpolarized incoming
electron beams as in (\ref{fbasym}). 
The RPV $s$-channel scalar exchanges in $e^+e^- \to t \bar t$ will
alter the SM FB asymmetry as follows:

\begin{eqnarray}
A_{FB(t)}^{SM} \to A_{FB(t)}^{RPV}=A_{FB(t)}^{SM} \times \left( 1+
\frac{\sigma_t^s}{\sigma_t^{SM}} \right)^{-1} \label{fbrpvtop}~.
\end{eqnarray}

\begin{figure}[htb]
\psfull
 \begin{center}
  \leavevmode
  \epsfig{file=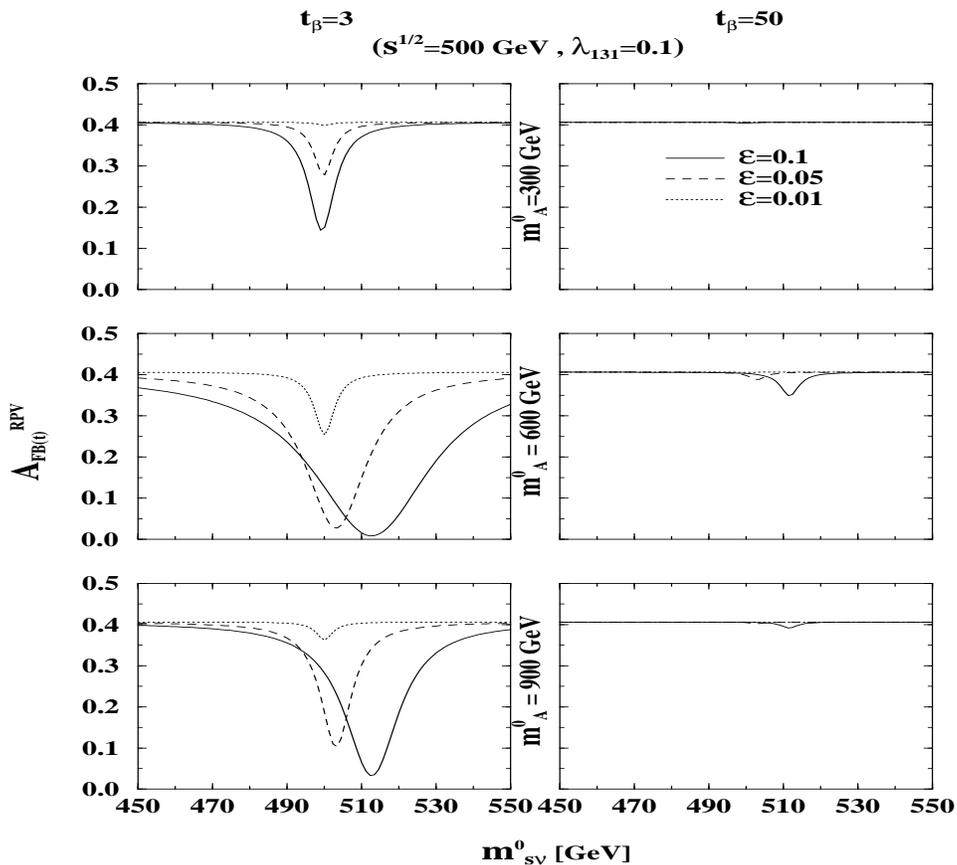,height=11cm,width=13cm,bbllx=0cm,bblly=1cm,bburx=20cm,bbury=26cm,angle=0}
 \end{center}
\caption{\emph{The FB asymmetry  $A_{FB(t)}^{RPV}$ defined
in (\ref{fbrpvtop}) as a function of
the RPC sneutrino mass $m_{s \nu}^0$ in an $e^+e^-$ collider
running at a c.m. energy of $\sqrt s=500$ GeV.
The asymmetry is shown for
$m_A^0=300,~600$ and $900$ GeV and for
$t_\beta=3$ (left figures) or $t_\beta=50$ (right figures).
For all combinations of $m_A^0$ and $t_\beta$,
$A_{FB(t)}^{RPV}$ is given for $\lambda_{131}=0.1$ and
for $\varepsilon=0.1,~0.05$ or $0.01$.
The corresponding SM value is $A_{FB(t)}^{SM} \sim 0.41$.}}
\label{fig8}
\end{figure}

\n In Fig.~\ref{fig8} we plot the FB asymmetry $A_{FB(t)}^{RPV}$ as
a function of the ``bare'' sneutrino mass $m_{s \nu}^0$
for a 500 GeV collider, for $t_\beta=3$ or $t_\beta=50$ with
$m_A^0=300,~600$ or 900 GeV. The FB asymmetry is shown for
$\lambda_{131}=0.1$ with three values
of the RPV scalar mixing parameter, $\varepsilon=0.1,~0.05$ and 0.01.
Only low $t_\beta$ values can give rise to
a significant shift from the SM FB asymmetry in $t \bar t$ production (the
very small shifts in the case of $t_\beta=50$ are bearly noticeable
on the scale used in Fig.~\ref{fig8}).
We see that these deviations from $A_{FB(t)}^{SM}$ can be
remarkable, reaching several tens of percent in a rather large range
of the sneutrino mass around 500 GeV.

We can, therefore, examine the expected limits on the
sneutrino--Higgs mixing scenario that
can be obtained purely from 
the FB asymmetry in $e^+e^- \to t \bar t$.
We again assume that
the central value of the
future measured FB asymmetry in $t \bar t$
production (at a c.m. energy of 500
GeV) takes the value of the corresponding SM
theoretical value, i.e., $A_{FB(t)}^{exp} = A_{FB(t)}^{SM}$, and that
the combined overall experimental error
is parametrized by the relative error $\delta_t^{FB}$ as:\footnote{Again
we assume that $\Delta A_{FB(t)}^{SM} \ll \Delta A_{FB(t)}^{exp}$. Otherwise
(\ref{deltatop}) should read: $\sqrt{(\Delta A_{FB(t)}^{exp})^2 +
(\Delta A_{FB(t)}^{SM})^2} \equiv  \delta_t^{FB} A_{FB(t)}^{exp} =
\delta_t^{FB} A_{FB(t)}^{SM}$.}

\begin{eqnarray}
\Delta A_{FB(t)}^{exp} \equiv \delta_t^{FB} A_{FB(t)}^{exp} =
\delta_t^{FB} A_{FB(t)}^{SM} \label{deltatop}~,
\end{eqnarray}

\n such that $\delta_t^{FB}=0.1$ implies a $10\%$ accuracy 
in $A_{FB(t)}^{exp}$.

Then, the 
deviation in the FB asymmetry due to
the RPV should be larger than the overall error
$\Delta A_{FB(t)}^{exp}$

\begin{eqnarray}
\mid A_{FB(t)}^{RPV} - A_{FB(t)}^{SM} \mid > \delta_t^{FB} A_{FB(t)}^{SM}
\label{fbcondition1}~.
\end{eqnarray}

\n In terms of the cross-sections (\ref{fbcondition1})
yields:

\begin{eqnarray}
\frac{\sigma_t^s}{\sigma_t^{SM}} >
\frac{\delta_t^{FB}}{1-\delta_t^{FB}} \label{fbcondition2}~.
\end{eqnarray}

Clearly, since $\delta_t^{FB}$ as well as $\delta_t^\sigma$ are
smaller than one, the condition in (\ref{fbcondition2}) is stronger
than the one obtained in (\ref{siglimittop}) through the
cross-section analysis when $\delta_t^{FB} = \delta_t^\sigma$.
However, as mentioned above, if the FB
asymmetry will be measured to a better accuracy than the
cross-section, i.e., $\delta_t^{FB} < \delta_t^\sigma$, and 
no deviation from the SM is detected, then
the limits obtained through (\ref{fbcondition2}) could be more
stringent than the ones obtained through (\ref{siglimittop}).

\begin{figure}[htb]
\psfull
 \begin{center}
  \leavevmode
  \epsfig{file=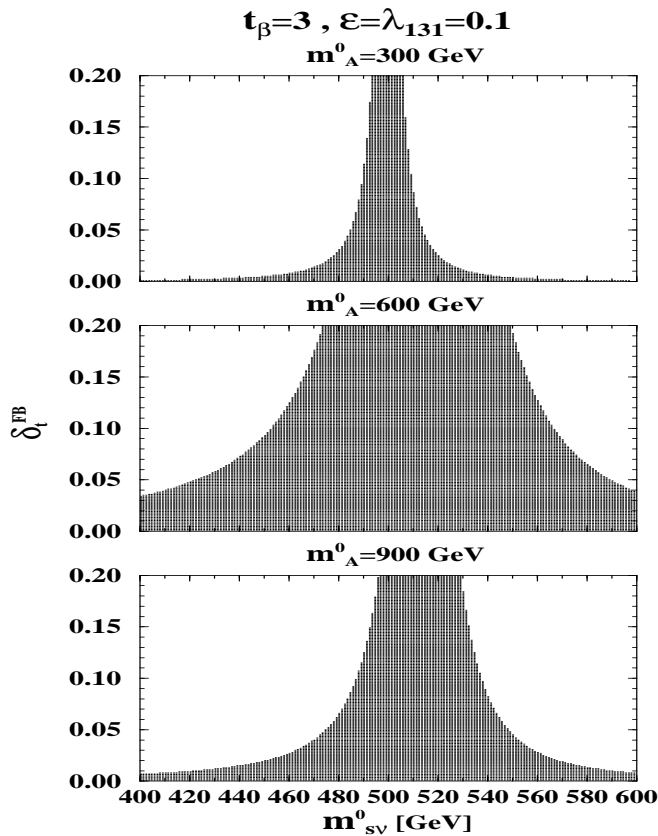,height=11cm,width=13cm,bbllx=0cm,bblly=1cm,bburx=20cm,bbury=26cm,angle=0}
 \end{center}
\caption{\emph{The shaded areas in the $\delta_t^{FB} - m_{s \nu}^0$ plane
represent values of $m_{s\nu}^0$ that can be
excluded for a given experimental error $\delta_t^{FB}$ in the actual
measurement of the FB asymmetry in $e^+e^- \to t \bar t$ at a 500 GeV
$e^+e^-$ collider. This is shown for $\varepsilon=\lambda_{131}=0.1$,
$t_\beta=3$ and for $m_A^0=300,~600$ or $900$ GeV.}}
\label{fig9}
\end{figure}

In Fig.~\ref{fig9} we show the sneutrino mass intervals that can be excluded
for a given $\delta_t^{FB}$ in
a measurement of the FB asymmetry in $e^+e^- \to t \bar t$
at a 500 GeV collider by requiring
(\ref{fbcondition2}).
The excluded mass intervals are plotted for the case
$\varepsilon=\lambda_{131}=0.1$, for $t_\beta=3$ and for
$m_A^0=300,~600$ or $900$ GeV.

Clearly, the FB asymmetry is a powerful probe of this signal or,
in case that no such signal is detected, is very useful in placing
limits  on this scenario. For example, if $A_{FB(t)}^{exp}$ is
measured at a 500 GeV machine with an error not exceeding $5\%$,
then a more than 100 GeV sneutrino mass interval can be excluded
if no deviation from the SM is observed for
$\varepsilon=\lambda_{131}=0.1$, $t_\beta=3$ and for   $m_A^0=600$
GeV. At the same time, a deviation in the
measured FB asymmetry will provide
a candidate signal for the $s$-channel scalar exchanges driven
by the sneutrino--Higgs mixing phenomena. Such additional signals
beyond just resonance enhancement in the cross-section should help
decipher the nature of the new physics involved. In particular, as
can be seen from (\ref{fbrpvtop}), a reduction (from the SM value)
in the FB asymmetry should give further evidence for $s$-channel
scalar exchanges which in turn should strengthen the theoretical
possibility of the sneutrino--Higgs mixing via the lepton number
violating RPVBT in (\ref{bterm}).

On the other hand, it should be noted that 
in the $ZZ/WW$ system the FB asymmetry is not as useful.
In particular, in $ZZ$ production 
$A_{FB(Z)}^{SM} = 0$ and,
therefore, also $A_{FB(Z)}^{RPV} = 0$ [see (\ref{fbrpvtop}) and 
replace the index $t$ with $Z$]. In such a case case no further
information can be gained from the FB asymmetry. The reaction
$e^+e^- \to W^+W^-$ does, however, have a non-zero FB asymmetry
within the SM and so the $s$-channel scalar exchanges will
decrease the FB asymmetry from $A_{FB(W)}^{SM}$ to $A_{FB(W)}^{RPV}$ 
according to
(\ref{fbrpvtop}). Unfortunately, due to the much larger SM $W^+W^-$
cross-section (as mentioned in the previous
section $\sigma_W^s/\sigma_W^{SM} \ll \sigma_Z^s/\sigma_Z^{SM}$),
the effect of the sneutrino-like resonance is too small to cause
an appreciable shift to $A_{FB(W)}^{SM}$ as long as our
dimensionless RPV parameters are kept below $\sim 0.1$. For
example, we find that, with $\varepsilon=\lambda_{131}=0.1$, in
the best cases (e.g., $m_A^0 =600$ GeV and $t_\beta=3$) the shift
$|A_{FB(W)}^{SM} - A_{FB(W)}^{RPV}|$ is at the level of a few
percent at most in a rather small $|m_{s \nu}^+ - \sqrt s|$
interval of several GeV.

\section{Summary and concluding remarks}

We have investigated a SUSY scenario in which lepton number is violated
in the scalar potential through a bilinear soft breaking term (RPVBT) 
as well
as in the superpotential through a Yukawa-like RPV trilinear operator 
(RPVTT).
The RPVBT gives rise to mixings between the Higgs and the slepton fields and
the new mass eigenstates of the neutral scalar sector
are sneutrino--Higgs admixtures.

We considered the case of small lepton number (or R-parity)
violation in the sense
that all lepton number violating couplings are typically at least an order of
magnitude smaller than their ``matching'' lepton number conserving
couplings in the R-parity conserving (RPC) SUSY Lagrangian.
In particular, we have used dimensionless R-parity violating (RPV)
couplings scaled to the
typical RPC couplings, and let these dimensionless couplings
be $\ll 1$.

We have carried out a detailed analysis of the
CP-even and CP-odd sneutrino--Higgs mass matrices in the presence
of a lepton number violating bilinear term in the SUSY scalar potential.
We have investigated their behavior under some limiting cases such as
small bilinear lepton number violating coupling, a heavy Higgs spectrum and
large $\tan^2 \beta$. Also, we have derived Feynman rules
for interaction vertices involving the new scalar mass-eigenstates
in the theory.

We then suggested that this small lepton number violating SUSY
framework may lead to new observable resonance formations in
scattering processes that are absent in RPC SUSY and in the
``usual'' RPV SUSY models in which scalar resonances can occur via
the trilinear Yukawa-like RPV interactions in the superpotential.
In particular, we focused on two particulary interesting channels,
$e^+e^- \to VV$, $V=Z$ or $W$ and $e^+e^- \to t \bar t$. The
resonance structure in these two channels arises only if there are
mixings between the sneutrino and Higgs states such that the
sneutrino component of the sneutrino--Higgs admixtures couple to
the incoming electron beams through Yukawa-like trilinear RPV 
interactions, while the Higgs component couples to the
massive gauge-bosons or to the top quarks. This makes the $VV$
and $t \bar t$ production channels unique as compared for example
to down-quark and charged lepton pair production in which the RPV
$s$-channel resonances occur via Yukawa-like trilinear RPV
interactions on both vertices. Such resonance signals in $e^+e^- \to VV$ and
$e^+e^- \to t \bar t$ may, therefore, serve as an efficient and
direct probe of the RPVBT in the SUSY scalar potential.

We found that the sneutrino-like scalars are expected to yield
a dominant resonance effect in both $e^+e^- \to VV$ and
$e^+e^- \to t \bar t$, and that the $t \bar t$ channel is much
more sensitive to the lepton number violating soft bilinear term.
Indeed, we have shown that
such a sneutrino-like resonance signal in $e^+e^- \to t \bar t$ is expected
to yield significant deviations in observables associated with top-quark pair
production which, under favorable circumstances,
can be detected in a 500 GeV $e^+e^-$ collider 
within a 100 GeV sneutrino mass
range around the c.m. energy,
either via ``simple'' event counting or via an analysis of
the Forward-Backward asymmetry.

If such a resonance will be observed in $e^+e^- \to t \bar t$,
then additional measurements should be carried out in the $VV$ production
channels as a cross-check for the existence of bilinear lepton number
violation in the SUSY scalar potential since
the later are expected to yield similar resonance signals.
The fact that the same resonance formation is expected to emanate
in two different scattering processes will help decipher the
nature of these resonance signals.

In the same vein, we have also considered the case that no such resonance
enhancement is or will be seen in existing and in future collider 
experiments.
First, we have used the recent LEP2 measurements of the $ZZ$ and $W^+W^-$
production cross-sections to place direct limits on the RPV SUSY parameters
involved in this scenario. Also, since this sneutrino--Higgs mixing effect
changes the theoretical prediction for the mass of the light SUSY Higgs
particle $h$, we have exploited the recent LEP2 bounds
on the $h$ mass to derive further limits on
the same lepton number violating SUSY parameters.
We found that the two independent LEP2
measurements are complementary for placing limits on these RPV parameters
and that they together exclude a significant portion of the
relevant SUSY parameter space involved.

In addition we have investigated the expected limits that can be placed
on this RPV SUSY scenario in a future 500 GeV $e^+e^-$ collider
in the case that no such resonance enhancement are detected
in $e^+e^- \to t \bar t$.

\bigskip
\bigskip
\bigskip

{\bf Acknowledgements:} G.E. thanks the
U.S.-Israel Binational Science Foundation, the Israel Science
Foundation and the Fund for Promotion of Research at the
Technion for partial support.


\begin{references}

\bibitem{rpreview} For reviews of R-parity violation see e.g.,
D.P. Roy, Pramana,\ J.\ Phys.\ {\bf 41}, S333 (1993); G.
Bhattacharyya, Nucl.\ Phys.\ {\bf B} (Proc. Suppl.) {\bf 52A}, 83
(1997); H. Dreiner, in {\it Perspectives in Supersymmetry}, edited
by G.L. Kane (World Scientific, Singapore, 1998), hep-ph/9707435;
P. Roy, hep-ph/9712520, Published in {\it Seoul 1997,
Pacific particle physics phenomenology}, 10-17,
talk given at APCTP Workshop:
Pacific Particle Physics Phenomenology (P4 97),
Seoul, Korea, 31 Oct - 2 Nov 1997.


\bibitem{rosiek} J. Rosiek, Phys.\ Rev.\ {\bf D41}, 3464 (1990),
unpublished Erratum in, hep-ph/9511250.

\bibitem{GH} We follow the notation in Y. Grossman and H.E. Haber,
Phys.\ Rev.\ {\bf D59}, 093008 (1999).

\bibitem{morebterms} S. Roy and B. Mukhopadhyaya,
Phys.\ Rev.\ {\bf D55}, 7020 (1997);
M.A. Diaz, J.C. Romao and J.W.F. Valle,
Nucl.\ Phys.\ {\bf B524}, 23 (1998);
C.-h. Chang and T.-f. Feng, hep-ph/9908295;
B. Mukhopadhyaya and S. Roy,
Phys.\ Rev.\ {\bf D60}, 115012 (1999).



\bibitem{davidson} S. Davidson, M. Losada and N. Rius, Nucl.\ Phys.\
{\bf B587}, 118 (2000).

\bibitem{snures} See e.g.,
S. Dimopoulos and L.J. Hall,
 Phys.\ Lett.\ {\bf B207}, 210 (1988);
V. Barger, G.F. Giudice and T. Han, Phys.\ Rev.\ {\bf D40}, 2987 (1989);
H. Dreiner and S. Lola,
in Munich/Annecy/Hamburg 1991, Proceedings of ``$e^+ e^-$
collisions at 500 GeV'', pages 707-711;
J. Erler, J.L. Feng and N. Polonsky, Phys.\ Rev.\ Lett.\ {\bf
78}, 3063 (1997);
J. Kalinowski {\it et al.}, Phys.\ Lett.\ {\bf
 B406}, 314 (1997);
M. Acciarri {\it et al.} (L3 collaboration), Phys.\ Lett.\ {\bf
B414}, 373 (1997);
S. Bar-Shalom, G. Eilam and A. Soni, Phys.\ Rev.\ Lett.\ {\bf 80},
4629 (1998).;
J.L. Feng, J.F. Gunion and T. Han, Phys.\ Rev.\ {\bf D58}, 071701 (1998);
D. Choudhury and S. Raychaudhury, hep-ph/9807373;
J. Kalinowski, hep-ph/9807312;
T.G. Rizzo, Phys.\ Rev.\ {\bf D59}, 113004 (1999);
T.G. Rizzo, hep-ph/9907344;
S. Lola , hep-ph/9912217;
G. Moreau, Ph.D. Thesis, hep-ph/0012156.

\bibitem{basis} J. Ferrandis, Phys.\ Rev.\ {\bf D60}, 095012
(1999).

\bibitem{neutrinomass} See e.g., H.-P. Nilles and N. Polonsky,
Nucl.\ Phys.\ {\bf B484}, 33 (1997);
B. Mukhopadhyaya and S. Roy,  Phys.\ Lett.\ {\bf
B443}, 191 (1998); M. Bisset {\it et al.},
Phys.\ Rev.\ {\bf D62}, 035001 (2000);
M. Hirsch {\it et al.}, Phys.\ Rev.\ {\bf D62}, 113008 (2000);

\bibitem{davidsonnew} S. Davidson and M. Losada, hep-ph/0010325.

\bibitem{GHnew} Y. Grossman and H.E. Haber, hep-ph/0005276.

\bibitem{rpvbtdec} See e.g., F. de Campos {\it et al.},
Nucl.\ Phys.\ {\bf B451}, 3 (1995); A. Akeroyd {\it et al.},
Nucl.\ Phys.\ {\bf B529}, 3 (1998).

\bibitem{hepph0005262} See e.g., K. Choi, E.J. Chun and K. Hwang,
J.\ Math.\ Phys.\ {\bf 42}, 130 (2001).

\bibitem{ourplb} S. Bar-Shalom, B. Mele and G. Eilam,
Phys.\ Lett.\ {\bf B500}, 297 (2001).

\bibitem{higgshunters} See e.g., J.F. Gunion, H.E. Haber, G. Kane
and S. Dawson, {\it The Higgs Hunter's Guide} (Addison-Wesley, New
York, 1990).


\bibitem{RPVeebb} See J. Erler, J.L. Feng and N. Polonsky in
Ref.\cite{snures}.

\bibitem{gunion} J.F. Gunion and H.E. Haber,
Nucl.\ Phys.\ {\bf B272}, 1 (1986), Erratum-{\it ibid.}, {\bf
B402}, 569 (1993).

\bibitem{djuadi} A. Djouadi {\it et al.}, Z.\ Phys.\ {\bf C74}, 93 (1997).

\bibitem{mhcorrections} S. Heinemeyer, W. Hollik and G. Weiglein,
hep-ph/0002213.

\bibitem{snumix} Y. Grossman and H.E. Haber, Phys.\
Rev.\ Lett.\ {\bf 78}, 3438 (1997); M. Hirsch, H.V. Klapor-Kleingrothaus and
S.G. Kovalenko, Phys.\ Lett.\ {\bf B398}, 311 (1997);
see also Refs.\cite{GH}, \cite{davidson} and \cite{davidsonnew}.

\bibitem{ourtautaupaper} See S. Bar-Shalom, G. Eilam and A. Soni in
Ref.\cite{snures}.

\bibitem{lep2mh} See e.g., talk given by Tom Junk at
the ``LEP Fest'', 10 October 2000, in
http://lephiggs.web.cern.ch/LEPHIGGS/talks.

\bibitem{ourcpsnu} S. Bar-Shalom, G. Eilam and A. Soni,
Phys.\ Rev.\ {\bf D59}, 055012 (1999).

\bibitem{lep2sigs} See summary notes and talks given by Salvatore Mele and
Anne Ealet in
http://lepww.web.cern.ch/lepww/talks\_notes. See also,
plots for the Summer 2000 conferences in
http://lepewwg.web.cern.ch/LEPEWWG/plots/summer2000.

\end{references}
\end{document}